\documentclass[twocolumn]{aastex701}

\usepackage{newtxtext,newtxmath}
\usepackage[T1]{fontenc}
\usepackage{graphicx}
\usepackage{amsmath}
\usepackage{hyperref}
\usepackage{longtable}
\usepackage[figuresright]{rotating}
\usepackage{multirow}
\usepackage{afterpage} 
\usepackage{enumitem}
\usepackage{placeins}

\begin{document}

\title{A High-Mass Size Deficit for Void Galaxies in the Complete SDSS DR7}

\author[0000-0002-0212-4563]{Olivia Curtis}
\affiliation{Department of Astronomy and Astrophysics, The Pennsylvania State University, 251 Pollock Road, University
Park, PA 16802, USA}
\affiliation{Institute for Gravitation and the Cosmos, The Pennsylvania State University, University Park, PA 16802, USA}
\affiliation{Penn State Extraterrestrial Intelligence Center, 525 Davey Laboratory, 251 Pollock Road, Penn State, University Park, PA, 16802, USA}
\affiliation{Department of Astronomy \& Institute for Astrophysical Research, 725 Commonwealth Ave., Boston University, Boston, MA 02215, USA}
\email[show]{ocurtis@psu.edu}

\author[0000-0001-6928-4345]{Bryanne McDonough}
\affiliation{Physics Department, Adelphi University, 1 South Ave., Garden City, NY 11530, USA}\affiliation{Department of Physics, Northeastern University, 360 Huntington Ave, Boston, MA 02115, USA}\affiliation{Department of Astronomy \& Institute for Astrophysical Research, 725 Commonwealth Ave., Boston University, Boston, MA 02215, USA}
\email{bmcdonough@adelphi.edu}

\author[0000-0001-7917-7623]{Tereasa G. Brainerd}
\affiliation{Department of Astronomy \& Institute for Astrophysical Research, 725 Commonwealth Ave., Boston University, Boston, MA 02215, USA}
\email{brainerd@bu.edu}

\begin{abstract}

We characterize how  environment shapes the sizes, luminosity functions, and mass functions of galaxies in large-scale underdensities by comparing the NSA catalog of the Sloan Digital Sky Survey Data Release 7 with an identically selected \texttt{TNG300} sample, both built with the void finding algorithm \texttt{VoidFinder}, and by assigning a local average underdensity contrast to each galaxy via a spherical top-hat smoothed density field approximation. Void galaxies have fainter characteristic magnitudes than their non-void counterparts, and, at fixed stellar mass, redshift, and color, NSA void galaxies with $M_*>10^{11}h^{-1}M_\odot$ are about $11\pm3\%$ more compact than galaxies in the field, a deficit the \texttt{TNG300} simulation reproduces. The galaxies that are responsible for this effect are almost all central and predominantly early type, i.e., the systems that are most likely to grow their outer envelopes via late-time mergers. These trends hold across two redshift bins to $z\leq0.114$. Crucially, this deficit is not a fixed property of the void sample but a steep function of the underdensity contrast, deepening toward the emptiest interiors and washing out at a typical density, implying that strict density control and sample selection are imperative when conducting environmental dependency studies like these. The same environmental dependence appears in the stellar mass function, which shifts to lower masses in the deepest voids, and in the close-pair (i.e., the ongoing merger) fraction, which falls toward the emptiest regions. Together, these point to a late-time, merger-driven growth of massive galaxies that is suppressed in voids, leaving their most massive members structurally distinct today.

\end{abstract}


\section{Introduction}
\label{sec:intro}

Cosmic voids are the largest and emptiest structures in the large-scale structure of the universe \citep{gregory1978}. Voids span $\sim10-100h^{-1}$Mpc (see, e.g., \citealt{sheth,sutter2012,sanchez2016cosmic}) and thus represent the largest way in which matter can organize itself in the universe. The interiors of cosmic voids are observed to be significantly underdense with respect to the background mean, regardless of the void size or the choice of tracer (\citealt{sanchez2016cosmic,mao2017,pollina2019,douglass2023}). Voids are useful cosmological probes and have well-characterized density profiles (\citealt{hamaus2014,pollina2017,curtis2022,schuster2023,curtis2025,Contarini2026}) that show flat, underdense interiors and steep ``ridges'' as their boundaries. Lastly, voids are dynamic and are influenced by the tidal forces of surrounding filaments and clusters (\citealt{bothun1992,tully2019};  \citealt{courtois2023,schuster2023}), allowing voids to accrete material over cosmic time \citep{vallesperez2021}.

For example, compared to halos and subhalos in more dense regions of the cosmic web, those within voids should form later in time (\citealt{liddle1993,sheth2004}). Recently, \cite{cavity1} and \cite{cavity2} have analyzed the star formation histories of local void galaxies to show that, compared to a sample of non-void field galaxies, void galaxies within the local universe assemble upwards of $50\%$ of their stellar mass up to $\sim1$~Gyr later. Within the context of the magnetohydrodynamical (MHD) simulation \texttt{TNG300} \citep{illustris2, illustris5, illustris1, illustris3, illustris4}, \cite{curtis2024} showed that void galaxies in the \texttt{TNG300} are marginally younger than non-void field galaxies, with luminosity-weighted stellar ages younger by $\sim150$ Myr.

Voids have been thought of as pristine environments for galaxy formation and evolution (\citealt{rojas2004,hoyle2012,kreckel2016,moorman2016}). For example, void galaxies are known to be relatively rich in neutral hydrogen (HI) for their luminosities (\citealt{kreckel2011,kreckel2012,beygu2013}), although recent results have found little difference between the gas properties of void and non-void galaxies \citep{Rodriguez2024}, and several authors have claimed that the lack of environment-driven quenching events impacts the chemical evolution of void dwarf galaxies (\citealt{hoeft2006,douglass2018,bidaran2025}). Lastly, it has been suggested that the fraction of active galactic nuclei (AGN) increases as the local matter density decreases (\citealt{kauffmann2004,ceccarelli2021,mishra2021,curtis2024,Aradhey2025,Curtis2026}, \citealt{Rouse2026}), possibly due to the fact that void galaxies are younger and have slower star formation histories than galaxies in filaments or clusters. However, void environments are still dynamically complex and several studies suggest that void environments may not be as pristine as traditionally thought. For example, minor mergers and local interactions between galaxies still shape void galaxy evolution \citep{Azevedo2026}, processed material from surrounding environments can accrete into voids \citep{vallesperez2021}, and void galaxies may undergo the same number of major mergers as those in filaments just at later times \citep{rodriguesmedrano2024}. 

The effects that a void environment has on the formation and evolution of galaxies are poorly constrained. It is clear from observational studies of void galaxies (e.g., \citealt{hoyle2004,Hoyle2005,Rojas2005,hoyle2012,kreckel2016,zaidouni2025,cavity1,cavity2}) that population-level studies are needed to truly understand how they differ from those in denser environments, especially if we are to constrain diverging evolutionary tracks across cosmic time (\citealt{cavity1,rodriguesmedrano2024,Curtis2026}). 

Void galaxy evolution also bears directly on the formation history of our own Galaxy. The Milky Way lies within the Local Void \citep{tully2019}, and its mass assembly is atypically quiet for a galaxy of its mass, having been built through only a handful of significant minor mergers and no major merger in recent cosmic time \citep{Lane2023,Evans2020}. This quiet accretion history is usually cast as a peculiarity of the Milky Way, yet it is precisely what one expects of a galaxy that assembled in a rarefied environment, where the merger rate is low and the late-time growth that mergers drive is correspondingly slow \citep{rodriguesmedrano2024}. If the Milky Way is, in this sense, a typical void galaxy, or, at least an inhabitant of a relatively diffuse sheet near the edge of the Local Void (e.g., \citealt{Neuzil2020}), then the population-level trends we measure here set the expectation for our own formation history, while the detailed, star-by-star data available for the Milky Way turn it into a nearby benchmark for how galaxies grow in the emptiest regions of the cosmic web. Placing the Galaxy in this context is a strong motivation for characterizing void galaxies as a population.

An open question is whether void galaxies are simply delayed in their evolution, offering a glimpse of an earlier epoch, or if they follow genuinely different tracks. The size-mass relation \citep{shen2003} is one test of this, but the literature disagrees on how environment imprints on it. Some studies find that void galaxies are more compact at fixed mass, most strongly at high mass \citep{Perez2025}, while others find that they are slightly larger \citep{Conrado2024}, and yet others conclude that they are unchanged once mass and morphology are fixed \citep{Porter2023,Abdullah2026}. Many do not control directly for the large-scale density contrast. Here, we show throughout that such a control is imperative and that spherical void finders are especially appropriate for this, because they return consistently deep underdensities (e.g., \citealt{douglass2023, curtis2024, Rincon2025}).

In this manuscript, we compare void galaxies in the updated Sloan Digital Sky Survey (SDSS; \citealt{SDSS}) Data Release 7 (DR7; \citealt{abazajian2009}) void catalog \citep{douglass2023} to a sample of galaxies that lie outside of voids. The SDSS remains one of the most spatially and spectroscopically complete galaxy samples available and thus remains a critical tool for understanding how the large-scale structure affects galaxy evolution. Indeed, several authors have already used the SDSS to study how the void galaxy colors, star formation histories, mass functions, sizes, and AGN fractions compare to galaxies in more dense environments (see, e.g., \citealt{rojas2004, Goldberg2005, Rojas2005, mishra2021, Rodriguez-Medrano2023, zaidouni2025, Aradhey2025}). 

\begin{figure}[!htbp]
    \centering
    \includegraphics[width=0.45\textwidth]{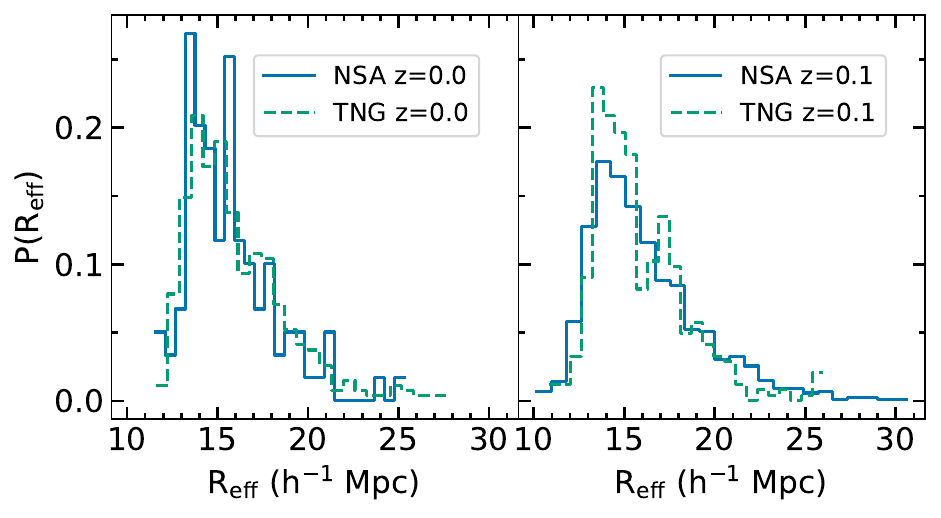}
    \caption{Probability density functions of void effective radii for voids in the \texttt{VoidFinder} catalogs of \cite{douglass2023} (solid blue) and the \texttt{TNG300} simulation (dashed green) at $z=0.0$ (left) and $z=0.1$ (right). The voids are identified from the $M_r\leq-20$ tracer catalog. The NSA and \texttt{TNG300} void catalogs span the same range of effective radii, so the observed and simulated void populations are closely matched.}
    \label{fig:ch7:Reff}
\end{figure}

Here, we aim to characterize the luminosity functions and mass-size relations of galaxies residing inside and outside of cosmic voids. Previous studies have examined these properties individually, with the luminosity function of void galaxies measured using an earlier SDSS release \citep{Hoyle2005}, while more recent work by \cite{Perez2025} identified modest differences in the mass-size relation for $\sim14{,}000$ galaxies in the SDSS DR16. In particular, they found that void galaxies tend to be smaller at fixed stellar mass than their counterparts in denser regions, an offset that they report across the mass range for early-type systems and that grows toward the high-mass end. In our work, we extend these previous efforts by leveraging the complete, volume-limited void catalog of \cite{douglass2023}. Since the void catalogs of \cite{douglass2023} are volume-limited samples that only extend to $z=0.114$ and depths of $M_r\lesssim-20.0$, our comparisons are restricted to the local universe. However, our results will form a basis for comparisons to next-generation observational surveys that will probe the universe to greater depths and, hence, higher redshifts.

Most directly, \cite{zaidouni2025} used this same \texttt{VoidFinder} catalog to show that void galaxies are bluer, fainter, and less massive than their non-void counterparts, trends that we recover here. Their study characterized the full flux-limited NASA-Sloan Atlas (NSA; \citealt{blanton2011}), a sample whose large faint blue population drives a strongly bimodal color distribution, whereas we work with the complete volume-limited $M_r\leq-20$ tracer sample so that void and non-void galaxies are always compared over an identical cosmic volume. We extend that work in three substantial respects. First, we measure the $r$-band luminosity functions and, for the first time using this catalog, the stellar mass-size relation of void galaxies, and we show that the most massive void galaxies are more compact than the field. Second, we build an identically selected \texttt{VoidFinder} sample in the \texttt{TNG300} simulation, so that every observed trend can be checked against a cosmological model that we analyze in exactly the same way. Third, and most importantly, we measure the average underdensity contrast of every host void and measure galaxy properties and ongoing merger history as a continuous function of that contrast, rather than through a single binary void vs. field label, which lets us show that the size deficit is specific to the deepest underdensities while ongoing mergers are more common in regions of typical matter density.

This manuscript is structured as follows. We discuss the galaxy catalogs and void finding algorithm in Section~\ref{sec:ch7:BOSS}. There, Section~\ref{sec:ch7:countsincells} maps the density field with counts in cells and Section~\ref{sec:ch7:morphmerg} summarizes the external morphology and merger information. In Section~\ref{sec:volumecalcs} we characterize the volumes and densities of the underdense environments that we study, while in Section~\ref{sec:ch7:lumfuncs}, we present luminosity functions for the galaxy populations. We then show stellar mass distributions in Section~\ref{sec:ch7:Ms} and size distributions in Section~\ref{sec:ch7:sizes}. In Section~\ref{sec:ch7:density}, we measure the galaxy properties as functions of the average underdensity contrast of each host void and compare the observed and simulated populations at fixed stellar mass. We present $(g-r)$ color distributions in Section~\ref{sec:ch7:galcolors}, galaxy morphologies in Section~\ref{sec:ch7:morphology}, and a recent-merger control in Section~\ref{sec:ch7:mergers}. We discuss these results in Section~\ref{sec:ch7:discussion}, where Section~\ref{sec:ch7:disc:trends} reviews the environmental trends, Section~\ref{sec:ch7:disc:contrast} shows that the size deficit tracks the underdensity contrast, Section~\ref{sec:ch7:disc:literature} reconciles the size-mass literature, Section~\ref{sec:ch7:disc:mergers} develops the suppressed merger-growth interpretation, and Section~\ref{sec:ch7:disc:sims} compares the observations with the \texttt{TNG300} simulation. We summarize our conclusions in Section~\ref{sec:ch7:conclusions}. Throughout, we adopt the \cite{Planck18} cosmology used to construct the void catalog of \cite{douglass2023} (i.e., $\Omega_{\Lambda,0} = 0.6889$, $\Omega_{m,0} = 0.3111$, $\Omega_{b,0} = 0.0490$, $\sigma_8 = 0.8102$, $n_s = 0.9665$, and $h = 0.6766$).

\begin{figure}[!htbp]
\centering
\includegraphics[width=0.46\textwidth]{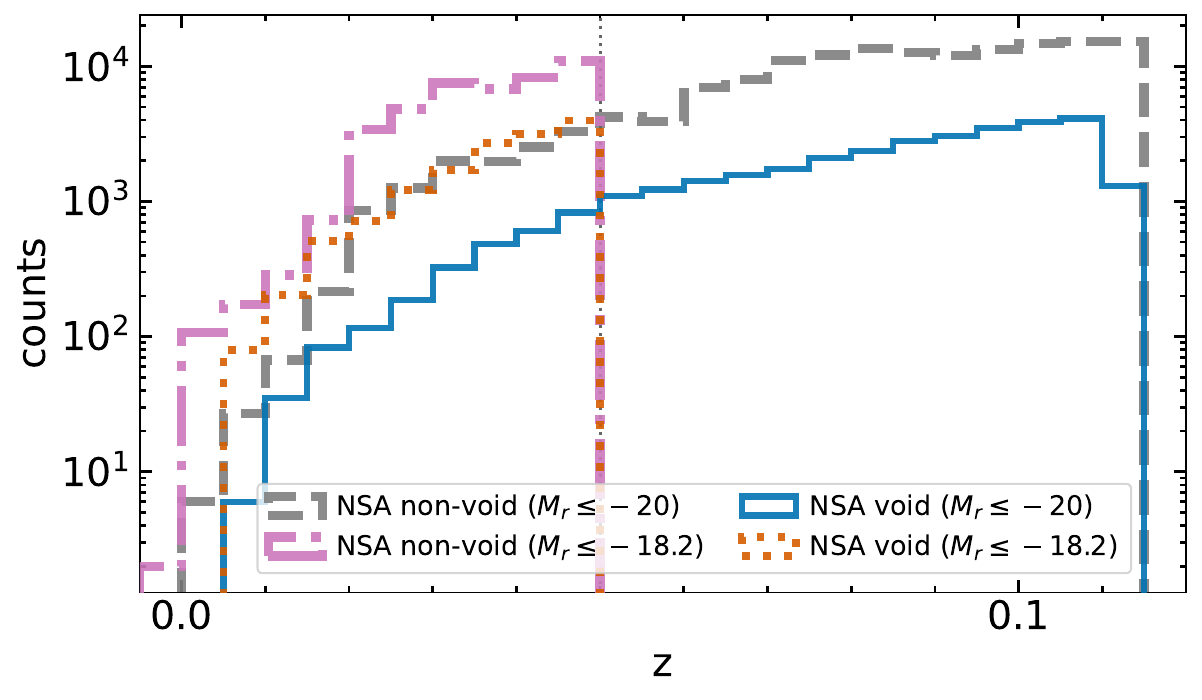}
    \caption{Redshift distributions for void (solid blue) and non-void (dashed gray) galaxies in the \texttt{VoidFinder} catalog of \cite{douglass2023}. All galaxies have $M_r\leq-20$. Void and non-void galaxies span the same redshift ranges, so any differences between them are not attributable to differences in redshift.}
    \label{fig:ch7:redshiftdist}
\end{figure}

\section{Methodology}
\label{sec:ch7:BOSS}

\begin{figure*}[!t]
    \centering
    \includegraphics[width=0.99\textwidth]{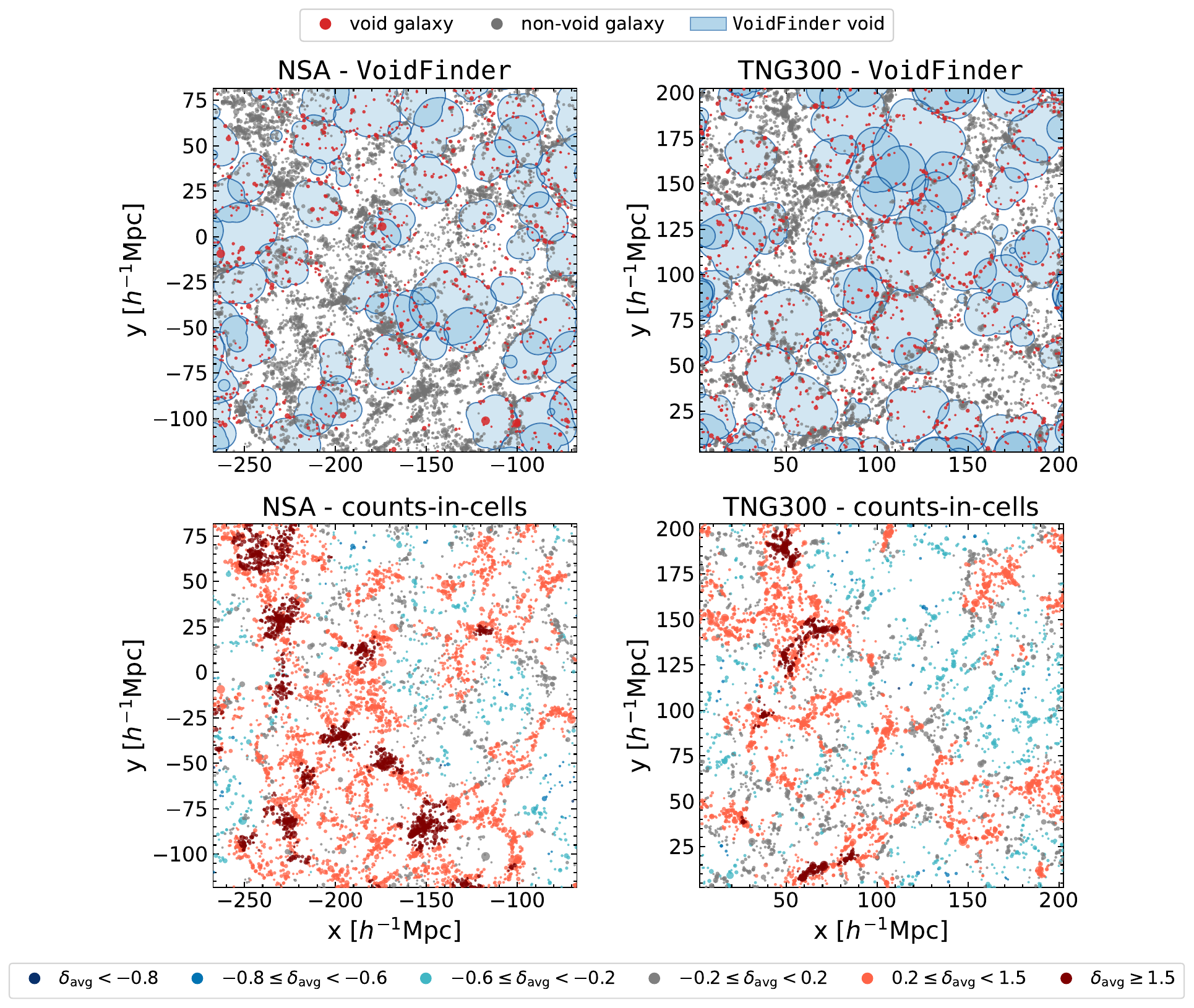}
    \caption{A $200\times200\times20\,h^{-1}$Mpc slice through the NSA sample (left column) and the \texttt{TNG300} simulation (right column), with every $M_r\leq-20$ galaxy drawn as a dot whose area scales with its $r$-band luminosity. Each row shows the two density-controlled large-scale structure classifications used throughout this manuscript. \textit{Top row:} the galaxies are colored by the \texttt{VoidFinder} void and non-void classification, and the \texttt{VoidFinder} voids that cross the slab are overplotted as the shaded regions, each one the union of its holes projected at the slab center. \textit{Bottom row:} each galaxy is colored by its corresponding counts-in-cells-derived underdensity contrast $\delta_{\rm avg}$ at its location, binned into six bins of $\delta_{\rm avg}$ such that the deepest voids appear in blue and the dense filaments and clusters in red. In each column the two rows trace the same large-scale structure.}
    \label{fig:ch7:slice}
\end{figure*}

We use the spherical void finder catalog of \cite{douglass2023} to obtain a sample of void galaxies in the observed universe. This catalog was produced using version 1.0.1 of the NSA catalog \citep{blanton2011}, which contains 641{,}409 galaxies observed in SDSS DR7 \citep{abazajian2009}. To construct a volume-limited survey, \cite{douglass2023} only select galaxies with absolute $r$-band magnitudes $M_r\leq-20$ and redshifts $z\leq0.114$. Spectroscopy was performed on all SDSS DR7 galaxies with Petrosian $r$-band magnitudes $m_r<17.77$ (\citealt{lupton2001,strauss2002}), and this redshift limit corresponds to the greatest distance that a galaxy with $M_r=-20$ was spectroscopically observed in SDSS DR7. As a result, the volume-limited sample of \cite{douglass2023} contains 194{,}125 galaxies. As will be discussed below, we bin this volume-limited sample into two sub-samples, finding void galaxies at redshifts below $z=0.05$ and void galaxies at redshifts $0.05 < z < 0.114$.

\cite{douglass2023} found voids in the NSA galaxy distribution using the \texttt{VoidFinder} algorithm of \cite{elad1997} and \cite{hoyle2004} as implemented in the publicly available Void Analysis Software Toolkit (\texttt{VAST}, \citealt{douglass2022}).\footnote{https://github.com/desi-ur/vast} In short, this algorithm identifies voids as a collection of overlapping spheres in a galaxy distribution. A visualization of this algorithm applied to both the NSA sample and the \texttt{TNG300}, alongside the counts-in-cells density field, can be seen in Figure~\ref{fig:ch7:slice} of Section~\ref{sec:ch7:countsincells}.

To run the algorithm, \cite{douglass2023} first converted galaxy sky coordinates and redshifts to  Cartesian coordinates via the transformation:

\begin{equation}
    x = s \cos(\alpha) \cos(\delta) \:,
\end{equation}

\begin{equation}
    y = s \sin(\alpha) \cos(\delta) \:,
\end{equation}

\noindent and

\begin{equation}
    z = s \sin(\delta) \: .
\end{equation}

\noindent Here, $\alpha$ is right ascension, $\delta$ is declination, and $s$ is the comoving distance to the galaxy. 

Each galaxy was then labeled as either a `wall galaxy' or an `isolated galaxy.' To do so, the distance to the third nearest neighbor of each galaxy was calculated along with the mean ($\mu$) and standard deviation ($\sigma$) of these values. `Isolated galaxies' were then defined as galaxies with distances to their third nearest neighbor greater than $\mu + 1.5\sigma$, and `wall galaxies' were those that were not labeled as `isolated.' Next, a 3D grid was drawn over the `wall galaxies,' and each empty cell was considered a potential void center. A sphere was then expanded around each empty cell until it reached a radius where it was bounded by four `wall galaxies.' To account for survey boundaries, \cite{douglass2023} rejected spheres if more than $10\%$ of their volume lay outside of the survey.

Spheres were then sorted by radius, and the largest sphere was labeled as the `maximal sphere' of the first void. The remaining spheres with radii greater than $10h^{-1}$Mpc were labeled as a `maximal sphere' if they did not overlap any previously defined `maximal sphere' by more than $10\%$ of their volume. Once all `maximal spheres' were defined, the remaining spheres that overlapped a `maximal sphere' by more than $50\%$ of their volume (i.e., spheres whose centers were within the `maximal sphere') were appended to the corresponding void region. 

Thus, each void contains a `maximal sphere' and a collection of smaller spheres that overlap it, resulting in voids with amorphous shapes. Each void is assigned an ``effective radius'' ($R_{\rm eff}$); i.e., the radius of a sphere that has the same volume as the void, defined as

\begin{equation}
    R_{eff} = \left(\frac{3V}{4\pi} \right)^{1/3} \; ,
\end{equation}

\noindent where $V$ is the volume of the void contained within the survey mask. In practice, spheres and voids are allowed some degree of overlap, so calculating void volumes is not trivial. Here, we calculate the volumes of the voids using a Monte Carlo approach, where we sampled the points within the volume of the survey and counted the fractions within each void compared to those within the survey. The \texttt{VoidFinder} catalog of \cite{douglass2023} contains 1{,}163 voids that comprise 39{,}735 spheres, and the voids have effective radii ranging from $10-31h^{-1}$Mpc. We present distributions of these effective radii in Figure~\ref{fig:ch7:Reff}. 

To create a catalog of void and non-void galaxies, we first converted the sky coordinates and redshifts of all NSA galaxies with $M_r\leq -20$ and $z\leq0.114$ to the same Cartesian coordinate system used by \cite{douglass2023}. %Following \cite{zaidouni2025}, we then define void galaxies as those that lie within or outside of any void and non-void galaxies as those that do not lie within a void. 
For our analysis in \S~\ref{sec:ch7:galprops}, we divided the galaxies into two redshift bins bounded by $z=0.0$, $0.05$, and $0.114$. Figure~\ref{fig:ch7:redshiftdist} shows the redshift distributions of the void galaxies that we obtained from the NSA catalog, and the vertical dashed line indicates the demarcation between the lower and higher redshift bins. The number of galaxies increases monotonically with redshift, such that there are 2{,}666 (12{,}234) void (non-void) galaxies in the lower redshift bin and 30{,}146 (143{,}014) void (non-void) galaxies in the higher redshift bin.

For comparison, we create similar \texttt{VoidFinder} catalogs using the $z=0.0$ and $z=0.1$ snapshots of the cosmological MHD galaxy formation simulation \texttt{TNG300}. The \texttt{TNG300-1} simulation (i.e., the highest-resolution public \texttt{TNG300} run) was propagated from $z=99$ to the present day and contains $2500^3$ dark matter particles and initial gas cells. With a gravitational softening length of $1.0\,h^{-1}\,\mathrm{kpc}$ at $z=0.0$ ($1.1\,h^{-1}\,\mathrm{kpc}$ at $z=0.1$), the simulation physically models the feedback from stellar and supermassive black hole outflows, radiative cooling from metal lines, the amplification of magnetic fields, as well as the chemical evolution and mass loss associated with evolving stellar populations.

We adopted the subhalo magnitudes from the supplementary catalog of \citet{illustris1} to assign luminosities to the galaxies. This supplementary catalog includes the effects of dust obscuration on the simulated galaxies and, thus, has magnitudes that better resemble SDSS magnitudes. Following \cite{illustris3}, when quoting stellar masses for our \texttt{TNG300} galaxies, we apply a correction to the stellar masses from the \texttt{TNG300} catalog (i.e., to account for the fact that the stellar masses in \texttt{TNG300} are not as well converged as in the benchmark \texttt{TNG100} simulation, as described in Appendix A of \citealt{illustris3}). Then, we select all \texttt{TNG300} subhalos that are cosmological in origin (i.e., those that have formed self-consistently via hierarchical clustering as opposed to numerical anomalies or halo fragments detected during halo identification), have $r$-band absolute magnitudes $M_r\leq -20$, and meet the galaxy size and mass resolution cuts that were imposed by \cite{Curtis2026}. These cuts require subhalos to be more massive than $10^{8.5}h^{-1}\rm{M_\odot}$ (such that each galaxy has at least $50$ stellar particles) and have radii larger than the softening radii of $[1+z]h^{-1}\rm{kpc}$. 

Next, we use \texttt{VAST} to run \texttt{VoidFinder} on these galaxies using the same parameters as described above. Figure~\ref{fig:ch7:Reff} shows the PDFs of void effective radii in the volume-limited catalog of \cite{douglass2023} (solid blue lines) and the \texttt{TNG300} simulation (dashed green) at $z=0.0$ (left) and $z=0.1$ (right). For the NSA catalog, a void is considered as part of the $z=0.0$ ($z=0.1$) sample if the redshift of its maximal sphere is $\leq0.05$ ($>0.05$). At $z=0.0$ ($z=0.1$) there are 108 (1{,}055) NSA voids and 415 (402) \texttt{TNG300} voids. The median effective radius for voids in the $z=0.0$ ($z=0.1$) NSA catalog is $15.36\pm0.27h^{-1}$Mpc ($15.49\pm0.11$), whereas the median effective radius for voids in the $z=0.0$ ($z=0.1$) snapshots of \texttt{TNG300} is $15.40\pm0.12h^{-1}$Mpc ($15.27\pm0.14h^{-1}$Mpc). %Thus, compared to the NSA catalogs of \cite{douglass2023}, VAST produces fewer voids, likely due to the comparatively smaller volume of the \texttt{TNG300}, of approximately the same size.

Finally, we create a sample of $46{,}841$ ($14{,}136$) void and $162{,}400$ ($69{,}209$) non-void galaxies in the $z=0.0$ ($z=0.1$) snapshot of the \texttt{TNG300} simulation, only selecting galaxies that meet the magnitude and resolution cuts described above. To match the deepest volume-complete NSA sample in each redshift bin, the $z=0.0$ snapshot is cut at $M_r\leq-18.2$ while the $z=0.1$ snapshot retains $M_r\leq-20$, and in both cases the voids themselves are defined by the $M_r\leq-20$ tracer catalog. We do not delineate between filament and cluster galaxies, so instead we focus our analysis on how the void environment shapes galaxy evolution. %For our \texttt{TNG300} galaxies, we have found that $\sim10\%$ ($\sim1\%$) of our non-void (void) galaxies exist within groups more massive than $10^{14}h^{-1}M_\odot$. Tests reveal that there are no major changes to our simulated results when we remove these galaxies from our sample of field galaxies. 

\subsection{Mapping the density field with counts in cells}
\label{sec:ch7:countsincells}

The \texttt{VoidFinder} catalog gives each galaxy a binary ``void'' or ``non-void'' label and, through its host void, an average underdensity contrast, $\delta_{\rm avg}$. To confirm that $\delta_{\rm avg}$ correlates to the trends that we report, and to extend the measurement to the full range of densities, we also measure it everywhere with a counts-in-cells estimate (i.e., the fixed-aperture approach of \citealt{croton2005}). To do so, we draw $6\times10^5$ random spheres of radius $10h^{-1}$Mpc, comparable to the mean void effective radius, throughout the NSA volume and the \texttt{TNG300} box. For the NSA sample we keep only spheres at least $90\%$ inside the survey mask and redshift shell following \cite{douglass2023}, and we compute the $\delta_{\rm avg}$ of each sphere from the same volume-limited $M_r\leq-20$ tracer catalog that defines the voids, so the sphere and void contrasts are measured identically. We then ascribe to each galaxy the average $\delta_{\rm avg}$ of all spheres that enclose it. Because the spheres sample volume rather than galaxies, this avoids the bias of a galaxy-centered aperture, which would always recover a near-average density. We briefly tested how our classification varies with both the number of spheres drawn and the sizes of the spheres, finding little change in our results at number counts and sizes larger than the ones that we used here.

Figure~\ref{fig:ch7:slice} shows a slice through both samples, showing the \texttt{VoidFinder} (top) and counts-in-cells $\delta_{\rm avg}$ (bottom) classification for the NSA (left) and \texttt{TNG300} (right) catalogs. From this figure, it is clear that the two classifications trace the same large-scale structure, because the \texttt{VoidFinder} voids coincide with the most underdense regions of the counts-in-cells map. The continuous $\delta_{\rm avg}$ therefore extends the binary void label smoothly across the full range of environments. Throughout our analysis below, we compare galaxies in dense and underdense regions of the cosmic web to those in regions of ``typical'' density (i.e., where $-0.2\leq\delta_{\rm avg}<0.2$).

\begin{figure}[!htbp]
    \centering
    \includegraphics[width=\linewidth]{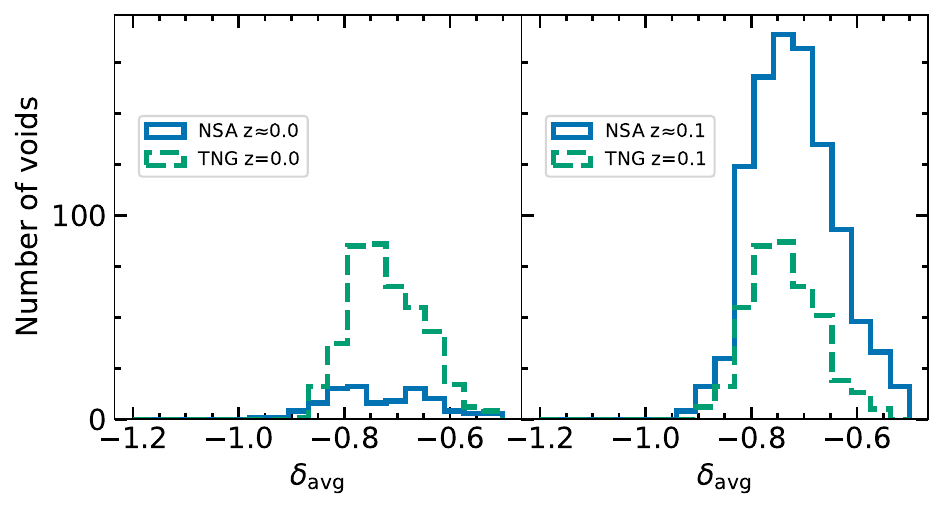}
    \caption{Distributions of void underdensity contrasts for NSA (blue) and \texttt{TNG300} (green) voids in the $z=0.0$ (left) and $z=0.1$ (right) samples. The voids are identified from the $M_r\leq-20$ tracer catalog. The NSA and \texttt{TNG300} voids reach the same characteristic underdensity, so the observed and simulated voids are equally empty.}
    \label{fig:voidUnderdensities}
\end{figure}

\subsection{Morphologies and recent mergers}
\label{sec:ch7:morphmerg}

Two of our tests draw on external morphology and merger information, which we summarize here. We use NSA morphologies from the Galaxy Zoo~1 project \citep{lintott2011}, in which volunteers visually classified nearly a million SDSS galaxies. We label a galaxy early type when its debiased elliptical vote fraction is $\gtrsim0.5$. For the \texttt{TNG300} simulation, we use the kinematic cold-disk fraction of \cite{genel2015}, calling a galaxy early type when fewer than $40\%$ of its stellar particles are in near-circular orbits.

\begin{deluxetable*}{lcccccc}[tbp]
\tablecaption{Survey, Void, and Field Volumes. Survey (or simulation) volumes, fraction of that volume that are \texttt{VoidFinder} voids, the total volume occupied by \texttt{VoidFinder} voids, the total volume of the field, the number of tracer galaxies, and the average number density of tracer galaxies. All volumes are comoving.\label{tab:VFSummary}}
\tablewidth{0pt}
\tablehead{
\colhead{Sample} &
\colhead{$V_{\rm survey}$} &
\colhead{$f_{\rm void}$} &
\colhead{$V_{\rm void}$} &
\colhead{$V_{\rm field}$} &
\colhead{$N_{\rm tracer}$} &
\colhead{$\bar{n}$} \\
&
\colhead{$(h^{-3}\,\mathrm{Mpc}^3)$} &
&
\colhead{$(h^{-3}\,\mathrm{Mpc}^3)$} &
\colhead{$(h^{-3}\,\mathrm{Mpc}^3)$} &
&
\colhead{$(h^{3}\,\mathrm{Mpc}^{-3})$}
}
\startdata
NSA: $0 < z \le 0.114$ & $2.73\times10^7$ & 0.609 & $1.66\times10^7$ & $1.07\times10^7$ & $188{,}060$ & $6.89\times10^{-3}$ \\
NSA: $0 < z \le 0.05$ & $2.41\times10^6$ & 0.589 & $1.42\times10^6$ & $9.91\times10^5$ & $14{,}900$ & $6.18\times10^{-3}$ \\
NSA: $0.05 < z \le 0.114$ & $2.49\times10^7$ & $0.606$ & $1.51\times10^7$ & $9.81\times10^6$ & $173{,}160$ & $6.97\times10^{-3}$ \\
\texttt{TNG300}: $z=0.0$ & $205^3$ & $0.581$ & $5.00\times10^6$ & $3.61\times10^6$ & $76{,}047$ & $8.83\times10^{-3}$ \\
\texttt{TNG300}: $z=0.1$ & $205^3$ & $0.552$ & $4.75\times10^6$  & $3.86\times10^6$ & $83{,}345$ & $9.67\times10^{-3}$ \\
\enddata
\end{deluxetable*}

This kinematic split is coarse, because it labels $95\%$ of the $M_r\leq-20$ \texttt{TNG300} sample early type, far from the Galaxy Zoo balance of $37\%$. We therefore also use the higher-resolution \texttt{TNG100} simulation for the morphology test alone. Deep-learning classifications of SDSS-like mock images exist only for \texttt{TNG100} \citep{huertascompany2019}, where they label $37\%$ of its $M_r\leq-20$ sample early type, the same balance as the Galaxy Zoo split. We use \texttt{TNG100} nowhere else because its $75\,h^{-1}$Mpc box contains only $15$ \texttt{VoidFinder} voids and about $500$ void galaxies, too few for any void-by-void measurement, whereas the \texttt{TNG300} box contains $415$ voids. The morphology test itself sets the environment by counts in cells rather than by the void catalog, so the small box still supports it to an extent.

We flag recent mergers by classifying all galaxies with an associated nearby companion. Here, a galaxy is said to have a companion if there is another galaxy within $50\,h^{-1}$kpc and $500$ km/s with a stellar mass between one quarter and four times its own, a standard selection calibrated against mergers \citep{patton2008,ventou2019}. We measure the fraction identically in the NSA sample from projected separations, and in the \texttt{TNG300} sample, from real-space separations. In both cases, the companions come from the same volume-limited $M_r\leq-20$ sample, so both members of every pair pass our fiducial selection, and we cross-check the NSA fractions against the visual merger flags that have been tabulated by citizen scientists through the Galaxy Zoo project \citep{darg2010}. We argue that, while the close-pair fraction traces the ongoing merger rate over the last few hundred Myr rather than the cumulative accreted mass, it is still a useful metric to quantify how galaxies in different environments are assembling mass on average at the times that they are observed. Several of our tests also separate central and satellite galaxies. We label a galaxy a central when it has no more massive $M_r\leq-20$ companion within $1\,h^{-1}$Mpc, measured in a projected cylinder with a $1000$ km/s velocity depth for the NSA sample and in a real-space sphere for the \texttt{TNG300} sample, and we label it a satellite otherwise.

\section{Results}
\label{sec:ch7:galprops}

\subsection{Void underdensity contrasts}
\label{sec:volumecalcs}

The average underdensity contrast of a void, $\delta_{\rm avg}$, can summarily characterize the bulk properties of its void galaxies \citep{Curtis2026}. In particular, \cite{Curtis2026} find that galaxies within the shell-crossing surface, where $\delta_{\rm avg}\sim-0.8$ to $-0.7$ \citep{blumenthal1992, sheth}, show the largest discrepancy compared to those in the field. We thus define the average underdensity contrasts of each void, defined as

\begin{equation}
    \delta_{\rm avg} = \frac{N_{\rm interior} /V}{\bar{n}} - 1 ,
\end{equation}

\noindent where $N_{\rm interior}$ is the number of galaxies that are members of a void with volume $V$ and $\bar{n}$ is the average number density of galaxies within the survey or simulation. In the case of the NSA voids, void volumes were defined as the union of all \texttt{VoidFinder} holes contained within the Hierarchical Equal Area iso-Latitude Pixelization (\texttt{HEALPix}; \citealt{Gorski2005}) survey mask \citep{douglass2023}. 

\begin{deluxetable*}{lcccc}[tbp]
\tabletypesize{\scriptsize}
\label{tab:ch7:lumfuncs}
\tablecaption{Single Schechter Fit Parameters}
\tablehead{
\colhead{Sample} &
\colhead{Environment} &
\colhead{$\phi^*$} &
\colhead{$M^*$} &
\colhead{$\alpha$} \\
&
&
\colhead{$(\mathrm{Mpc}^{-3}\,h^3)$} &
&
}
\startdata
\footnotesize
NSA: $z\leq0.05$ & Void
& $(7.13 \pm 0.29) \times 10^{-3}$ & $-20.170 \pm 0.037$ & $-1.010 \pm 0.026$ \\
                   & Non-void
& $(2.54 \pm 0.05) \times 10^{-2}$ & $-20.604 \pm 0.019$ & $-0.982 \pm 0.012$ \\
\hline
NSA: $0.05<z\leq0.114$ & Void
& $(5.06 \pm 0.11) \times 10^{-3}$ & $-20.399 \pm 0.029$ & $-0.794 \pm 0.044$ \\
                      & Non-void
& $(2.47 \pm 0.03) \times 10^{-2}$ & $-20.807 \pm 0.013$ & $-0.934 \pm 0.016$ \\
\hline
\texttt{TNG300}: $z=0.0$ & Void
& $(7.13 \pm 0.11) \times 10^{-3}$ & $-20.351 \pm 0.016$ & $-0.802 \pm 0.012$ \\
                         & Non-void
& $(2.95 \pm 0.03) \times 10^{-2}$ & $-20.733 \pm 0.009$ & $-0.806 \pm 0.006$ \\
\hline
\texttt{TNG300}: $z=0.1$ & Void
& $(7.97 \pm 0.17) \times 10^{-3}$ & $-19.951 \pm 0.034$ & $0.092 \pm 0.075$ \\
                         & Non-void
& $(3.27 \pm 0.05) \times 10^{-2}$ & $-20.727 \pm 0.018$ & $-0.703 \pm 0.024$ \\
\enddata
\end{deluxetable*}

We compute an effective survey comoving volume by first calculating the full-sky comoving volume between $z=0.0$ to $z=0.114$, sampling $10{,}000{,}000$ random points within this 3D volume, and counting the fraction of points that fall within the survey mask ($f_{\rm sky}$). The survey volume is then defined as

\begin{equation}
    V_{\rm survey} = f_{\rm sky} \frac{4}{3} \pi (r_{\rm max}^3 - r_{\rm min}^3) \; ,
\end{equation}

\noindent where $r_{\rm max}$ ($r_{\rm min}$) is the comoving distance at $z=0.114$ and ($z=0.0$). 

Void volumes were calculated in a similar fashion. Each Monte Carlo point was queried against its nearest void such that the fraction of points inside voids, $f_{\rm void}$, estimates the fraction of the survey volume that is located within a \texttt{VoidFinder} hole. The total volume within voids is then calculated as

\begin{equation}
    V_{\rm void} = f_{\rm void} V_{\rm survey} 
\end{equation}

\noindent such that the non-void (i.e., field) volume becomes 

\begin{equation}
    V_{\rm field} = V_{\rm survey} - V_{\rm void} \:.
\end{equation}

\begin{figure}[!htbp]
    \centering
    \includegraphics[width=0.47\textwidth]{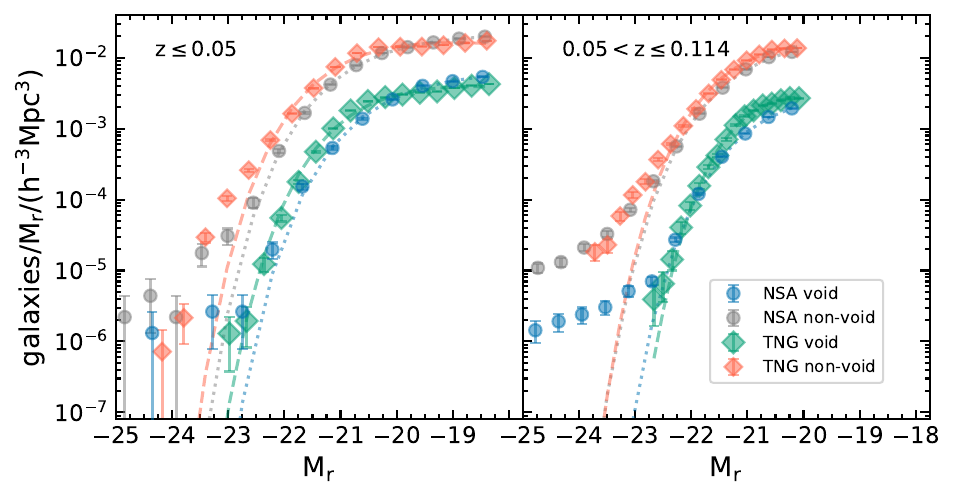}
    \caption{Absolute $r$-band luminosity functions of NSA void (blue circles), NSA non-void (gray circles), \texttt{TNG300} void (green diamonds), and \texttt{TNG300} non-void (red diamonds) galaxies at $z=0.0$ (left) and $z=0.1$ (right). The dashed lines show best-fit Schechter luminosity functions. Error bars show the standard errors of the mean distributions. The $z=0.0$ panel uses the deeper $M_r\leq-18.2$ volume-complete sample and the $z=0.1$ panel uses $M_r\leq-20$. Within each redshift bin the void galaxies have a fainter characteristic magnitude than the field in both samples.}
    \label{fig:ch7:lumfuncs}
\end{figure}

\begin{figure*}[tbp]
    \centering
    \includegraphics[width=0.75\textwidth]{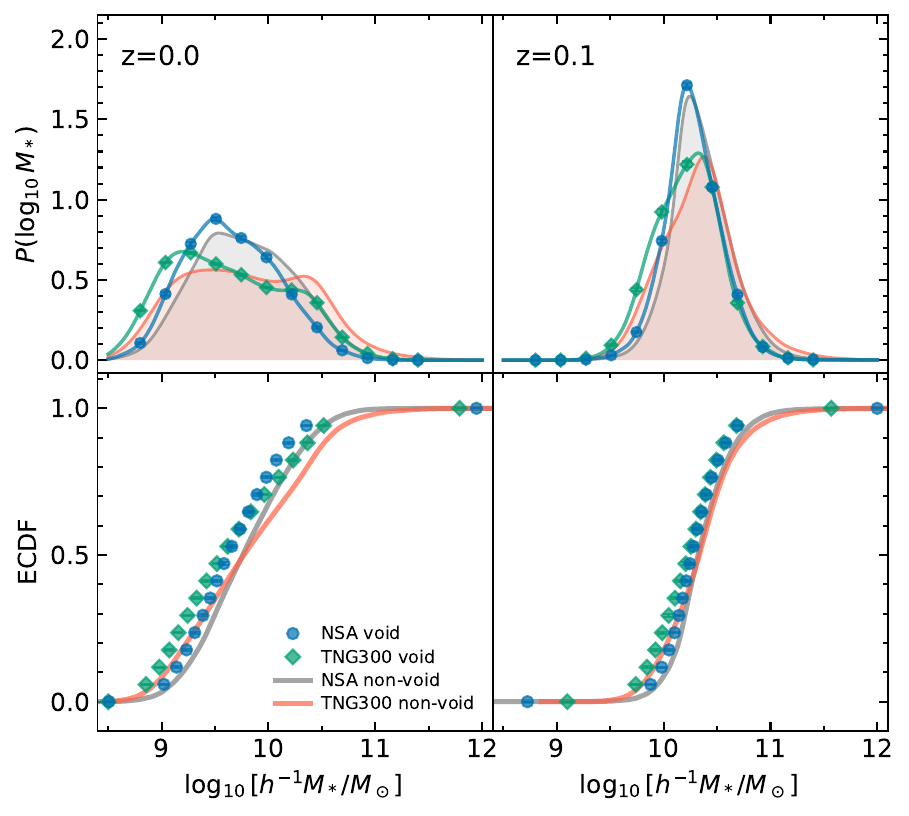}
    \caption{\textit{Top:} kernel density estimates of the stellar masses of NSA void (blue circles), NSA non-void (shaded gray), \texttt{TNG300} void (green diamonds), and \texttt{TNG300} non-void (shaded red) galaxies, with the void populations drawn as points and the non-void populations as shaded densities. \textit{Bottom:} the empirical cumulative distribution functions of the same populations, with the non-void populations drawn as thick lines. Each row shows the $z=0.0$ (left) and $z=0.1$ (right) samples, and the error bars and bands show bootstrap standard errors. The $z=0.0$ panels use the deeper $M_r\leq-18.2$ volume-complete sample and the $z=0.1$ panels use $M_r\leq-20$. The void galaxies are shifted toward lower stellar masses than the field in both the NSA data and the simulation.}
    \label{fig:ch7:Mscdf}
\end{figure*}

\noindent We present the survey (and simulation) volumes, void covering fractions, void volumes, field volumes, and average tracer number densities in Table~\ref{tab:VFSummary}. For the NSA sample, there were $188{,}060$ total tracer galaxies, and we calculate $f_{\rm sky}=0.177$. We then repeat these calculations in our two redshift bins, taking into account the fact that the redshift boundary can divide a \texttt{VoidFinder} void. In this case, only the fraction of the voids (and the galaxies within them) that lie within a given redshift cut are used in volume and density calculations. Lastly, we repeat these calculations for the $z=0.0$ and $z=0.1$ snapshots of the \texttt{TNG300} simulation. In general, compared to the corresponding \texttt{TNG300} samples, the low- and high-redshift NSA catalogs have similar values of $f_{\rm void}$ and slightly lower values of $\bar{n}$, the latter of which can explain why our \texttt{TNG300} voids tend to be slightly smaller on average (see, e.g., \citealt{Sutter2014b, curtis2025}). 

We use these volumes to calculate the average underdensity contrasts of our voids in Figure~\ref{fig:voidUnderdensities}. For the NSA sample, $\delta_{\rm avg}$ was calculated before applying any redshift cuts. For the purposes of constructing the histograms, an NSA void is considered to be part of the $z=0.0$ ($z=0.1$) sample if the center of its maximal sphere had a redshift $z\leq0.05$ ($z>0.05)$. We find good agreement between the average underdensity contrasts between the NSA sample and the \texttt{TNG300}. At $z=0.0$ ($z=0.1$) the NSA voids have median values of $\delta_{\rm avg}=-0.712\pm0.022$ ($\delta_{\rm avg}=-0.722\pm0.003$) and the \texttt{TNG300} voids have median values of $\delta_{\rm avg}=-0.726\pm0.004$ ($\delta_{\rm avg}=-0.741\pm0.004$). We thus conclude that our void samples are very well characterized and are accurately probing the deeply underdense regime near the shell-crossing threshold. This deep and consistent underdensity is precisely what our measurement requires. We show in Section~\ref{sec:ch7:density} that the high-mass size deficit is specific to genuinely underdense regions and washes out toward a typical density, so the tightly clustered deep contrasts that \texttt{VoidFinder} returns are what make the deficit measurable, which motivates the strict control on $\delta_{\rm avg}$ that we adopt throughout.

\subsection{Luminosity Functions}
\label{sec:ch7:lumfuncs}

For the luminosity function, stellar mass, and color comparisons that follow, we use the deepest volume-complete sample in each redshift bin, $M_r\leq-18.2$ at $z\leq0.05$ and $M_r\leq-20$ at $0.05<z\leq0.114$. The size-mass relation and the density-binned residuals of Sections~\ref{sec:ch7:sizes} and~\ref{sec:ch7:density} retain the $M_r\leq-20$ sample as the size deficit is concentrated at the high-mass end where the deepening adds no galaxies. The one exception is the overall size distributions at the start of Section~\ref{sec:ch7:sizes}, which use the deeper $z\leq0.05$ sample. The voids themselves are always identified from the $M_r\leq-20$ tracers, taken from the catalog of \cite{douglass2023} for the NSA sample and from our own \texttt{VoidFinder} runs for the \texttt{TNG300}, so the deepening changes only which galaxies we examine, not the voids. Wherever applicable, each figure caption states which sample was used in its creation.

Figure~\ref{fig:ch7:lumfuncs} shows absolute $r$-band luminosity functions of void and non-void galaxies in the $z=0.0$ (left) and $z=0.1$ (right) samples. The dashed lines show best-fit Schechter luminosity functions \citep{schechter1976}. The observed luminosity functions are well described by a single Schechter function over the full magnitude range, whereas the deeper $z\leq0.05$ sample reveals that the simulated luminosity functions depart from a single Schechter form, showing a faint-end excess that is most pronounced for the \texttt{TNG300} void galaxies together with an excess of the brightest galaxies relative to the exponential cutoff. We nonetheless adopt a single Schechter function throughout since a double-Schechter parameterization is not robustly favored by the observed samples.%This is more noticeable for the NSA galaxies in the higher redshift bin, which flattens at the brightest magnitudes. 

Table~\ref{tab:ch7:lumfuncs} shows the parameters for the best-fit Schechter luminosity functions. Because the $z\leq0.05$ bin reaches the deeper $M_r\leq-18.2$ limit while the $0.05<z\leq0.114$ bin is limited to $M_r\leq-20$, the fits in the two bins span different luminosity ranges and are not directly comparable between redshifts. The deeper low-redshift fits, which sample well beyond the knee of the luminosity function, recover steeper faint-end slopes than the brighter-limited high-redshift fits. However, within each redshift bin, compared to the void galaxies, the non-void galaxies have systematically brighter characteristic magnitudes (e.g., for NSA galaxies in the higher redshift bin, $M_{*,\rm{non-void}}=-20.807\pm0.013$ and $M_{*,\rm{void}}=-20.399\pm0.029$). There are no consistent differences between the faint-end slopes of the void and non-void galaxy populations.

Compared to the simulated void galaxies, the observed void galaxies have values of $M_*$ that are $0.181\pm0.040$ magnitudes fainter in the lower redshift bin, where the fit extends to the deeper $M_r\leq-18.2$ limit, and $0.449\pm0.045$ magnitudes brighter in the higher redshift bin where the fits are less well constrained due to the volume-complete magnitude limit. Similarly, the best-fit Schechter functions have values of $\Delta \alpha \equiv \alpha_{TNG300}-\alpha_{NSA}$ of $0.208\pm0.029$ ($0.886\pm0.087$) at $z=0.0$ ($z=0.1$), indicating that the simulated void galaxies have consistently flatter slopes at dimmer magnitudes. Compared to the observed non-void galaxies, the simulated non-void galaxies also have brighter values of $M_*$ and shallower faint-end slopes, but the differences are less pronounced. For example, in the higher redshift bin, $M_{*, \rm{non-void}}^{\rm NSA}=-20.807\pm0.013$ but $M_{*, \rm{non-void}}^{\texttt{TNG300}}=-20.727\pm0.018$. %Lastly, the faint ends of the luminosity functions of \texttt{TNG300} galaxies asymptotically approach slightly higher values of $\phi_*$, suggesting the fraction of the volume of \texttt{TNG300} that is occupied by voids is slightly lower than it is in the NSA catalog. %Thus, while the best-fit Schechter parameters do not change significantly between redshift bins, there are differences between the luminosity functions of simulated vs. observed void galaxies.  

\subsection{Stellar Masses}
\label{sec:ch7:Ms}

The top row of Figure~\ref{fig:ch7:Mscdf} shows the stellar mass probability density functions (PDFs) for NSA void (circles), NSA non-void (shaded gray), \texttt{TNG300} void (diamonds), and \texttt{TNG300} non-void (shaded red) galaxies at $z=0.0$ (left) and $z=0.1$ (right). Compared to the non-void galaxies, there is a slight overabundance of void galaxies with stellar masses between $10^{9.5}-10^{10}h^{-1}M_\odot$ and a slight underabundance of void galaxies with stellar masses $M_*\gtrsim10^{10.5}h^{-1}\rm{M_\odot}$. Although there are no differences between the peaks of the void and non-void distributions, at $z=0.0$, where the deeper $M_r\leq-18.2$ sample reaches the low-mass population, the simulated distributions peak at lower stellar masses than the corresponding NSA distributions. For instance, the NSA galaxies reach a peak probability density near $M_*=0.3\times10^{10}h^{-1}M_\odot$ while the \texttt{TNG300} galaxies peak at even lower stellar masses, near $0.15$ to $0.25\times10^{10}h^{-1}M_\odot$, both well below the $M_r\leq-20$ peaks of $\sim1.6\times10^{10}$ and $\sim2.3\times10^{10}h^{-1}M_\odot$ recovered in the $0.05<z\leq0.114$ bin.

\begin{figure}[!htbp]
    \centering
    \includegraphics[width=0.8\columnwidth]{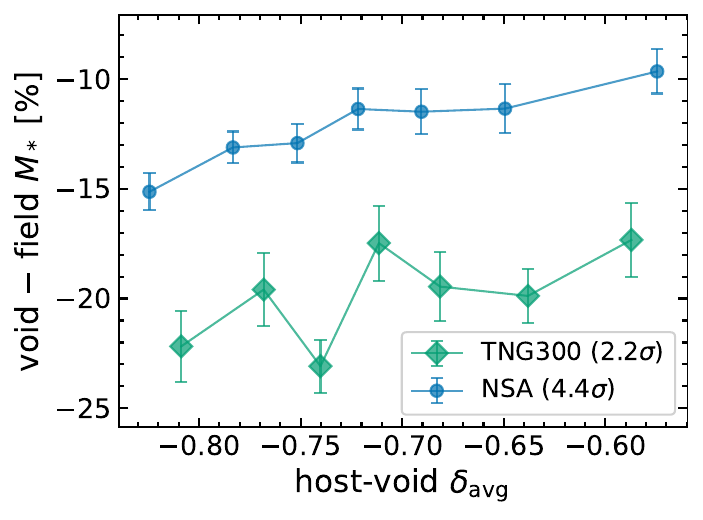}
    \includegraphics[width=0.8\columnwidth]{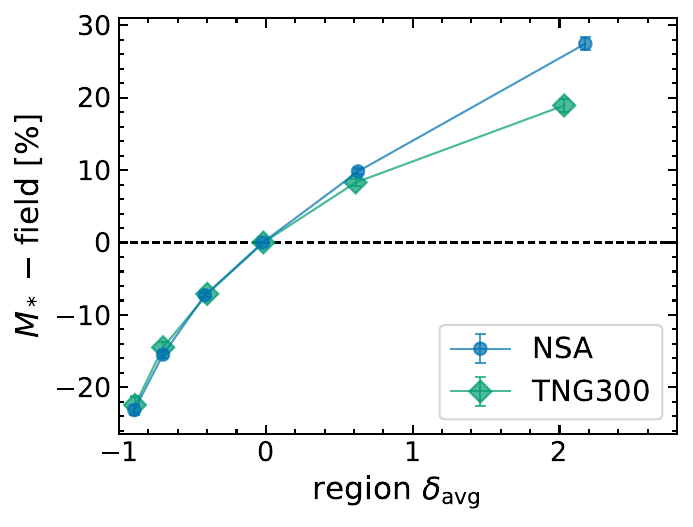}
    \caption{\textit{Top:} the void-minus-field stellar mass residual (in percent) as a function of host-void $\delta_{\rm avg}$, for the NSA (blue circles) and \texttt{TNG300} (green diamonds). The legend gives the significance of a linear trend with $\delta_{\rm avg}$, and the error bars are computed by resampling whole voids. \textit{Bottom:} the median stellar mass relative to the typical-density galaxies as a function of the counts-in-cells underdensity contrast, $\delta_{\rm avg}$, for the NSA and \texttt{TNG300} central galaxies. In both views, the stellar mass function shifts steadily with environment such that the most underdense regions host the least massive populations. Here all galaxies have $M_r\leq-20$.}
    \label{fig:ch7:massdensity}
\end{figure}

\begin{figure}[!htbp]
    \centering
    \includegraphics[width=\columnwidth]{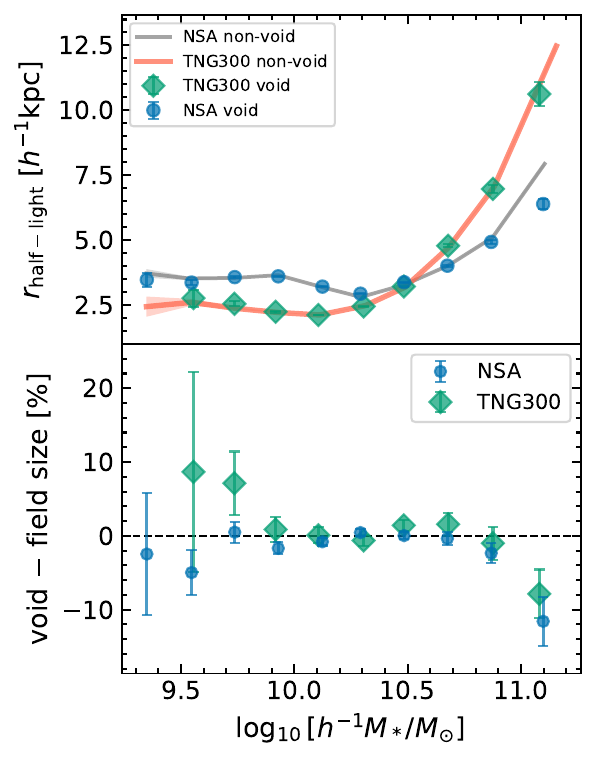}
    \caption{\textit{Top:} the global median half-light radius as a function of stellar mass for the NSA void (blue circles), NSA non-void (gray line and band), \texttt{TNG300} void (green diamonds), and \texttt{TNG300} non-void (red line and band) galaxies, which is the global size-mass trend with no environmental control applied. \textit{Bottom:} the same comparison after controlling for stellar mass, redshift, and color, our matched void-minus-field size residual in percent, in which each NSA void galaxy is measured against the field galaxies of its own stellar mass, redshift, and $(g-r)$ color, and each \texttt{TNG300} void galaxy against the field of its own stellar mass. The most massive void galaxies, $M_*>10^{11}h^{-1}M_\odot$, are about $11\%$ (NSA) and $7\%$ (\texttt{TNG300}) more compact than the matched field, while at lower masses the residual is consistent with zero. All galaxies have $M_r\leq-20$. Error bars and bands resample whole voids on the void side and bootstrap the field.}
    \label{fig:ch7:sizemass}
    \label{fig:ch7:residual}
\end{figure}

The bottom row of Figure~\ref{fig:ch7:Mscdf} shows empirical cumulative distribution functions (ECDFs) for these galaxy samples. Compared to the non-void galaxies, the void galaxies tend to be slightly less massive on average. At $z=0.0$ ($z=0.1$), the median stellar mass for the NSA void galaxies is $(0.42\pm0.01)\times10^{10}h^{-1}M_\odot$ (($1.83\pm0.01)\times10^{10}h^{-1}M_\odot$) while the median stellar mass for the NSA non-void galaxies is $(0.57\pm0.01)\times10^{10}h^{-1}M_\odot$ ($(2.10\pm0.01)\times10^{10}h^{-1}M_\odot$), which corresponds to differences of $\sim0.14$ dex at $z=0.0$, where the deeper $M_r\leq-18.2$ sample reaches lower masses, and $\sim0.06$ dex at $z=0.1$. As was the case before, the differences between the simulated void and non-void galaxy distributions are slightly more pronounced. For example, at $z=0.1$, the median stellar mass for the simulated void galaxies is $(1.70\pm0.01)\times10^{10}h^{-1}M_\odot$ while the simulated median stellar mass for the simulated non-void galaxies is $(2.16\pm0.01)\times10^{10}h^{-1}M_\odot$.  %Compared to the observed galaxies, the simulated galaxies span a slightly wider range of stellar masses but otherwise follow similar trends. The median stellar masses for the NSA (TNG300) void galaxies are $1.77^{+0.02}_{-0.02}\times10^{10}\:h^{-1}M_\odot$ ($1.89^{+0.02}_{-0.02}\times10^{10}\:h^{-1}M_\odot$) at $z=0.0$ and $1.79^{+0.01}_{-0.01}\times10^{10}\:h^{-1}M_\odot$ ($1.68^{+0.01}_{-0.02}\times10^{10}\:h^{-1}M_\odot$) at $z=0.1$, indicating that there is little difference between the median stellar masses of observed and simulated void galaxies.

\begin{figure*}[!tbp]
    \centering
    \includegraphics[width=0.75\textwidth]{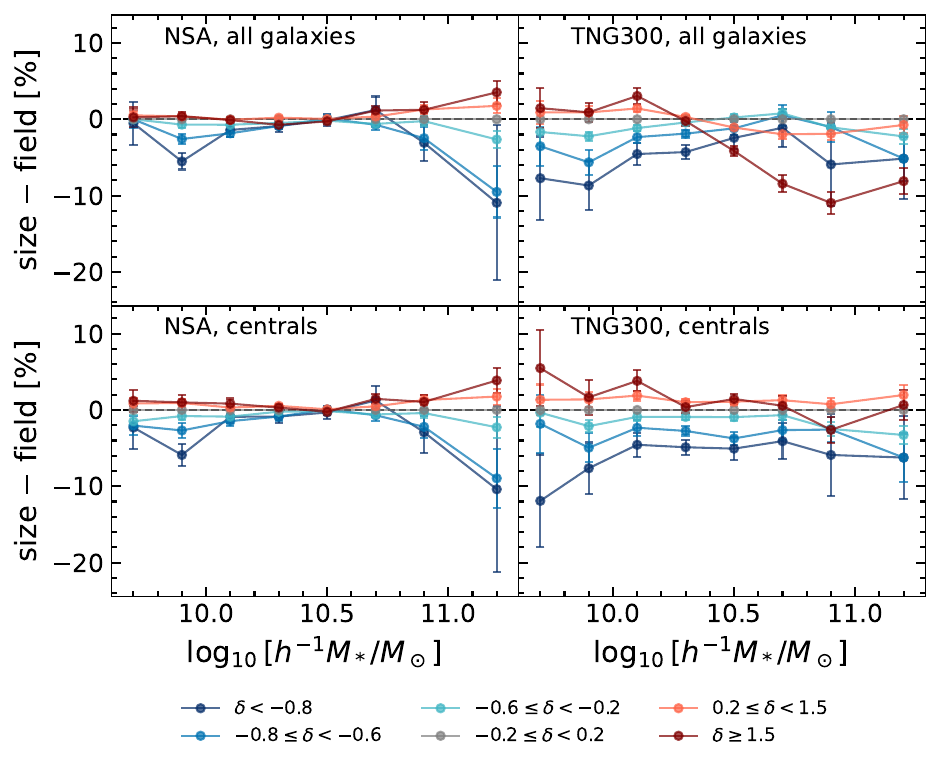}
    \caption{Half-light radius relative to the typical-density galaxies ($-0.2\leq\delta_{\rm avg}<0.2$), in six bins of the local large-scale underdensity contrast $\delta_{\rm avg}$, for the NSA sample (left) and the \texttt{TNG300} (right). Each NSA galaxy is compared to the field galaxies of the same stellar mass, redshift, and $(g-r)$ color, and each \texttt{TNG300} galaxy to the field galaxies of the same stellar mass, so the residual is measured at fixed mass, redshift, and color in the data and at fixed mass in the simulation. \textit{Top row:} all $M_r\leq-20$ galaxies. \textit{Bottom row:} central galaxies only. The high-mass size deficit deepens monotonically toward more underdense environments and reverses to an excess in overdense regions. For all galaxies, the densest \texttt{TNG300} bins instead turn downward owing to the simulation's tidally stripped cluster satellites, an effect that is removed in the centrals-only row, where the NSA and \texttt{TNG300} trends agree across the full range of $\delta_{\rm avg}$. As in Figure~\ref{fig:ch7:sizemass}, the error bars on the void points are computed by resampling whole voids rather than individual galaxies, so they reflect the void-to-void variance in each bin.}
    \label{fig:ch7:randsmr}
\end{figure*}

The binary void label from the \texttt{VoidFinder} algorithm can be sharpened into a continuous label that reflects the void underdensity. Since we measured $\delta_{\rm avg}$ for every host void in Section~\ref{sec:volumecalcs} and across the whole density field in Section~\ref{sec:ch7:countsincells}, we can ask how the stellar mass function depends on the depth of the underdensity, controlling each void galaxy against an average region of the field (i.e., galaxies with an associated underdensity contrast of $-0.2\leq\delta_{\rm avg}<0.2$) at the same redshift. Figure~\ref{fig:ch7:massdensity} shows the results for both of these methods, with the top panel showing the void-minus-field stellar mass residual as a function of host-void $\delta_{\rm avg}$. In the NSA sample, the mass deficit deepens steadily toward more underdense interiors, from about $-9.7\%$ in the shallowest voids to about $-15\%$ in the deepest, a trend that is significant at $4.4\sigma$ from the seven-bin weighted fit and at $4.2\sigma$ from a bin-free per-galaxy regression, implying the result does not depend on the binning. \texttt{TNG300} reproduces the sign and slope at lower significance, $2.2\sigma$, over the narrow range of $\delta_{\rm avg}$ its voids span and with a larger offset.

The bottom panel extends the measurement across the full range of density with the counts-in-cells $\delta_{\rm avg}$ of Section~\ref{sec:ch7:countsincells}. In both samples, the median stellar mass falls smoothly toward the deepest underdensities and rises in overdense regions, and the \texttt{VoidFinder} interiors sit at the underdense end of this relation. Because the mass function itself depends on environment, every size and color comparison that follows is made at fixed stellar mass since not doing so would usually reveal void galaxies to be smaller as a population compared to field galaxies due to the fact that the former tend to be less massive on average. 

\subsection{Sizes}
\label{sec:ch7:sizes}

The NSA half-light radii are $r$-band elliptical Petrosian radii, while the \texttt{TNG300} radii are derived from the supplementary catalog of \cite{genel2018}. To determine the radii of the \texttt{TNG300} galaxies, \cite{genel2018} first selected all subhalos that contained more than 100 stellar particles and created projections of them along the $z$-axis of the simulation. Then, they constructed $r$-band surface density maps using the stellar particles bound to that subhalo. The half-light radii were then defined as the radii that enclose half of the light in the surface brightness maps. As these catalogs were only made for the $z=0.0$, $0.5$, $1.0$, $1.5$, $2.0$, $3.0$, $4.0$, and $5.0$ catalogs, we can only compare the half-light radii of galaxies in the lower redshift bin. At $z=0.0$, there were 44{,}318 void galaxies with half-light radii larger than $1h^{-1}$kpc.

In the NSA sample, without fixing color or redshift as we do below, the void and non-void galaxies appear to have seemingly identical size distributions, with median half-light radii of $2.45\pm0.01h^{-1}$kpc and $2.43\pm0.01h^{-1}$kpc, respectively. The \texttt{TNG300} void galaxies are slightly smaller on average, $2.10\pm0.01h^{-1}$kpc vs.\ $2.27\pm0.01h^{-1}$kpc for the field, and both simulated medians fall below the observed values. The simulated distribution has a slightly heavier tail toward the largest radii, with $4.0\%$ of the \texttt{TNG300} void galaxies exceeding $6h^{-1}$kpc vs.\ $1.3\%$ of the NSA void galaxies. %Thus, compared to the \texttt{TNG300} void galaxies, the NSA void galaxies tend to have $r$-band half-light radii that are larger by a factor of $1.29\pm0.07$ (corresponding to a difference of $0.68\pm0.17h^{-1}$kpc). 

The top panel of Figure~\ref{fig:ch7:sizemass} shows this global trend in median half-light radius against stellar mass. All four populations follow a rising relation, and the void and non-void galaxies track one another closely below $\sim10^{10.5}h^{-1}M_\odot$ where the NSA galaxies are somewhat larger than their \texttt{TNG300} counterparts. At the highest masses, $M_*\gtrsim10^{11}h^{-1}M_\odot$, the NSA void galaxies fall below the field. The bottom panel quantifies this with a residual comparing each galaxy to the field at its own mass, color, and redshift. That is, for each galaxy $i$ we form

\begin{equation}
r_i = \log_{10} r_{{\rm half},i} - \log_{10}\tilde{r}_{\rm field}\!\left(M_{*,i},(g-r)_i,z_i\right),
\label{eq:ch7:resid}
\end{equation}

\noindent where $r_{{\rm half},i}$ is its half-light radius and $\tilde{r}_{\rm field}$ is the median half-light radius of the non-void galaxies at the same stellar mass $M_{*,i}$, color $(g-r)_i$, and redshift $z_i$, read from a grid fit to the field. We control all three quantities for the NSA sample, and stellar mass alone for the \texttt{TNG300}, whose single snapshot has no redshift range and whose colors are too blue and narrow for an appropriate field reference to be drawn. Despite being a volume-limited survey, we do this calculation at fixed redshift since the observable of interest, an $r$-band half-light radius, is more affected by surface brightness dimming, atmospheric distortion, etc., each of which can cause galaxies further away to appear smaller at these distances. We ultimately report the deficit in a mass bin as the difference $\mathrm{median}(r_{\rm void})-\mathrm{median}(r_{\rm field})$, where $\mathrm{median}(r_{\rm void})$ is the median of this residual taken over the void galaxies in the bin and $\mathrm{median}(r_{\rm field})$ the same median taken over the non-void galaxies, the latter close to zero by construction. Independent tests indeed reveal that the difference is unbiased and returns zero when applied to two random halves of the field.

The most massive void galaxies (i.e., those with $M_*>10^{11}h^{-1}M_\odot$) are thus $11\pm3\%$ more compact than the matched field in the NSA sample, at $3.6\sigma$, and $7\pm3\%$ in the \texttt{TNG300}, at $2.3\sigma$. Controlling for color brings the NSA deficit down from about $16\%$ at fixed mass and redshift alone, and we adopt the fully controlled value as the more conservative estimate. At lower masses the residual is consistent with zero in both samples, so the compactness is a property of the rarest, most massive void galaxies alone.

The deficit is not a measurement systematic. It is unchanged when $r$-band luminosity replaces the photometric stellar mass, when galaxy concentration is added to the controls, and when the NSA sample is restricted to galaxies well above the $1.4''$ seeing. A mock void sample drawn from the field alone, mass-matched but carrying no environmental effect, returns a residual consistent with zero. In the following subsections, we show how this relationship evolves as a function of void underdensity as well as under other constraints (e.g., at fixed morphology or color group). 

\begin{figure*}[!tbp]
    \centering
    \includegraphics[width=0.99\textwidth]{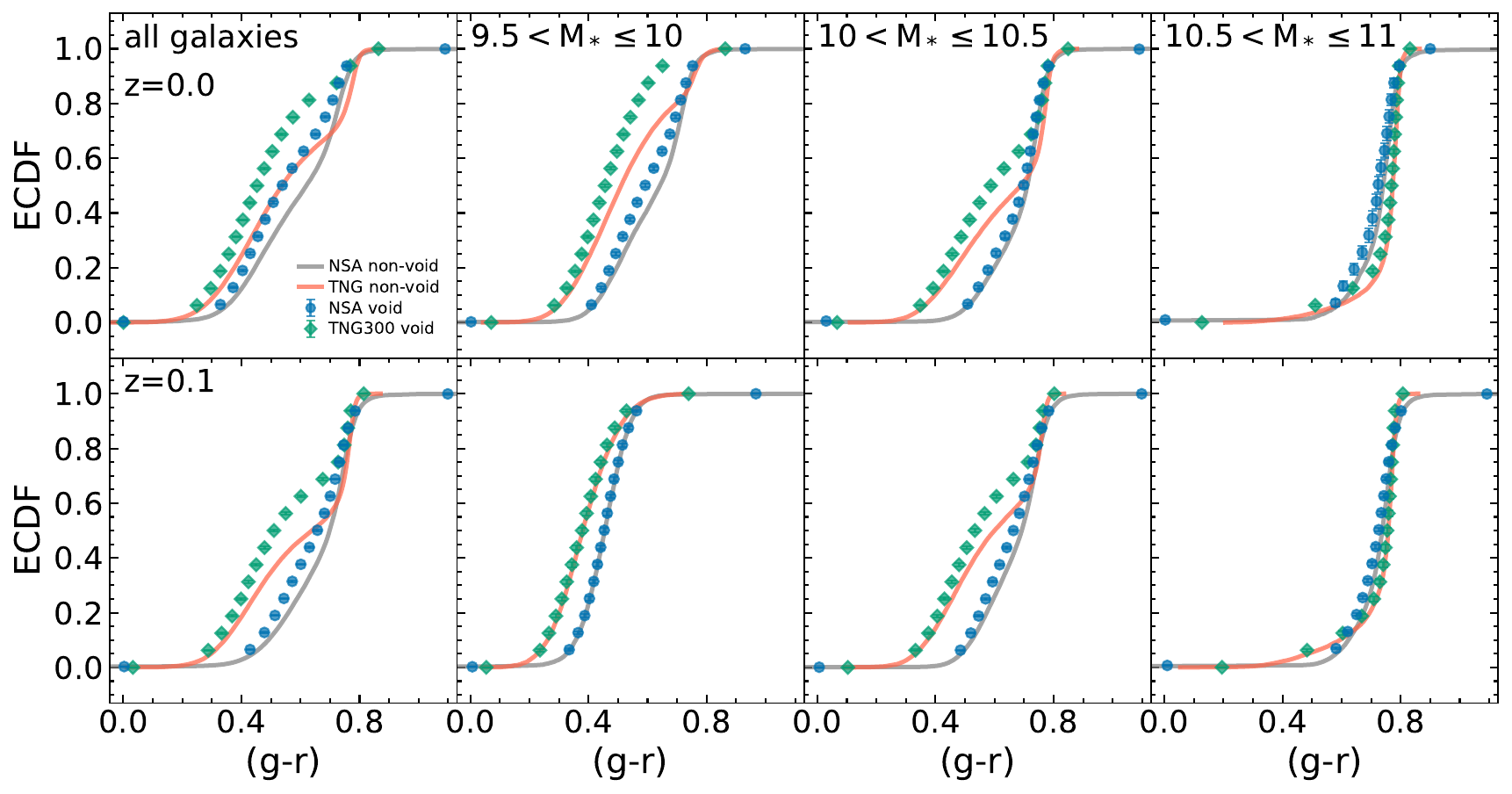}
    \caption{Empirical cumulative distribution functions of $(g-r)$ color for NSA void (blue circles), NSA non-void (shaded gray), \texttt{TNG300} void (green diamonds), and \texttt{TNG300} non-void (shaded red) galaxies. The top row is the $z=0.0$ sample and the bottom row is the $z=0.1$ sample. The leftmost column uses all galaxies, and the next three columns use the stellar mass bins $10^{9.5}$ to $10^{10.0}$, $10^{10.0}$ to $10^{10.5}$, and $10^{10.5}$ to $10^{11.0}h^{-1}M_\odot$ from left to right. Following the deepening described in Section~\ref{sec:ch7:lumfuncs}, the $z=0.0$ row uses the deeper $M_r\leq-18.2$ volume-complete sample while the $z=0.1$ row uses $M_r\leq-20$. The void galaxies are bluer than the field in every mass bin, most clearly at intermediate stellar mass.}
    \label{fig:ch7:colorcdf}
\end{figure*}

\subsubsection{Sizes as a function of void underdensity}
\label{sec:ch7:density}

\begin{deluxetable}{lcc}[!htbp]
\tabletypesize{\footnotesize}
\caption{Satellite fraction of the $M_*>10^{10.5}h^{-1}M_\odot$ galaxies in each bin of large-scale underdensity contrast, weighted by the random spheres as in Figure~\ref{fig:ch7:randsmr}.}
\label{tab:ch7:satfrac}
\tablehead{\colhead{$\delta_{\rm avg}$ bin} & \colhead{NSA} & \colhead{\texttt{TNG300}}}
\startdata
$\delta_{\rm avg}<-0.8$ & 0.08 & 0.04 \\
$-0.8\leq\delta_{\rm avg}<-0.6$ & 0.11 & 0.08 \\
$-0.6\leq\delta_{\rm avg}<-0.2$ & 0.17 & 0.14 \\
$-0.2\leq\delta_{\rm avg}<0.2$ & 0.23 & 0.21 \\
$0.2\leq\delta_{\rm avg}<1.5$ & 0.31 & 0.31 \\
$\delta_{\rm avg}\geq1.5$ & 0.44 & 0.47 \\
\enddata
\end{deluxetable}

The size deficit is not a fixed property of the void population but, rather, it is a steep function of the large-scale density. Figure~\ref{fig:ch7:randsmr} shows the fixed-mass residual in six bins of the counts-in-cells $\delta_{\rm avg}$ of Section~\ref{sec:ch7:countsincells}. In each mass bin we reference the residual of Equation~\ref{eq:ch7:resid} to the typical-density bin (i.e., $-0.2\leq\delta_{\rm avg}<0.2$) so that bin is zero by construction and every other bin measures the offset from typical-density galaxies of the same mass, redshift, and color. The high-mass deficit reaches about $10\%$ in the NSA sample and $6\%$ in the \texttt{TNG300} in the deepest underdensities, weakens toward higher density, and reverses to a small excess in the densest regions, of the kind reported for cluster galaxies \citep{yoon2017}. The deepest bins match the environments of our \texttt{VoidFinder} voids and recover the same amplitude as the matched void deficit. Averaged over all masses, the residual is flat with $\delta_{\rm avg}$, at $0.2\sigma$, and the $(g-r)$ color at fixed mass shows no trend either, at $1.1\sigma$. The environmental imprint on galaxy structure is therefore confined to the most massive galaxies, particularly those in the emptiest regions.

The \texttt{TNG300} trend turns down at the densest end, which is a satellite effect since massive satellites are tidally stripped and are unusually compact in the simulation. It is absent among the central galaxies and does not affect the void comparison, as voids are populated almost entirely by host galaxies, whose satellite fractions match closely between the two samples (Table~\ref{tab:ch7:satfrac}). We note that we are unable to demarcate central and satellite galaxies in our void vs.\ non-void experiment due to the small number of satellite galaxies in \texttt{VoidFinder} voids.

\subsection{Colors}
\label{sec:ch7:galcolors}

The high-mass size deficit raises the question of whether it reflects the differing color mixtures of the two environments rather than a genuine structural difference, since void galaxies are bluer and galaxy size at fixed mass depends on color \citep{Abdullah2026}. We first characterize the color distributions, then test the high-mass size deficit at fixed color.

\begin{figure}[!htbp]
    \centering
    \includegraphics[width=\columnwidth]{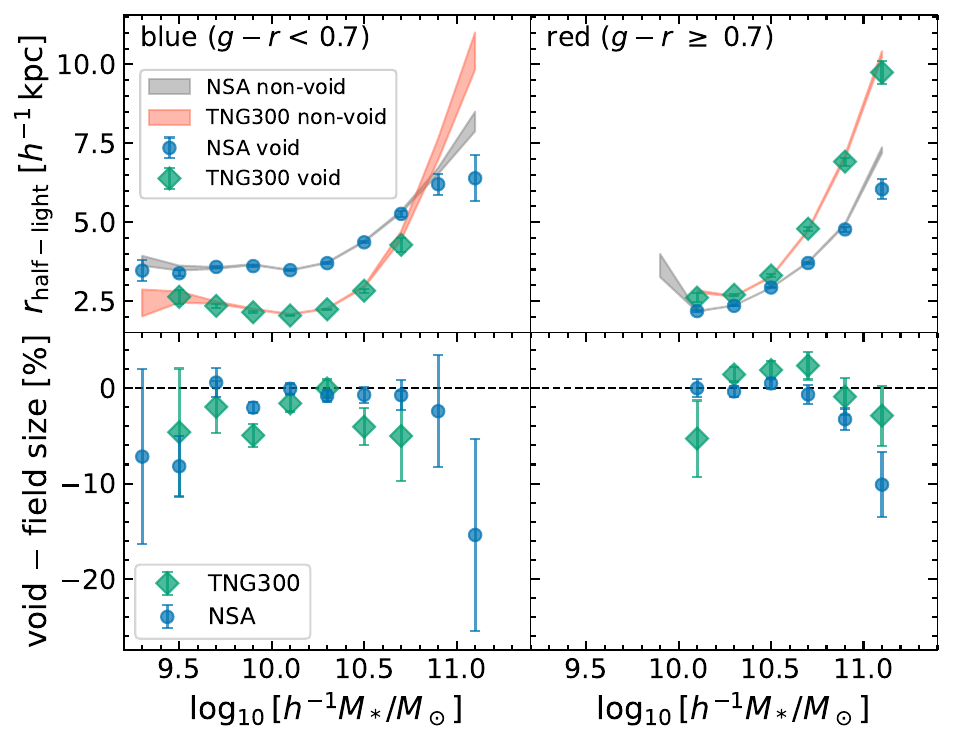}
    \caption{Half-light radius as a function of stellar mass split by color. \textit{Left column:} the blue galaxies ($g-r<0.7$). \textit{Right column:} the red galaxies ($g-r\geq0.7$). \textit{Top row:} the median half-light radius of the void galaxies, drawn as NSA blue circles and \texttt{TNG300} green diamonds, against the non-void median shown as the shaded gray (NSA) and red (\texttt{TNG300}) bands. \textit{Bottom row:} the void-minus-field size residual in percent, measured within each color against the field galaxies of that same color, at fixed stellar mass, redshift, and color for the NSA sample and at fixed stellar mass for the \texttt{TNG300}, with the NSA error bars from resampling whole voids and the \texttt{TNG300} error bars at the galaxy level. At the highest masses the void galaxies are smaller than the fixed-color field in both colors and in both samples, so the deficit is not produced by the differing color mixtures of the two environments. The blue \texttt{TNG300} void relation stops near $10^{10.7}h^{-1}M_\odot$ because massive blue galaxies are rare in the simulated voids, where the bright void population is almost entirely red. All galaxies have $M_r\leq-20$.}
    \label{fig:ch7:sizecolor}
\end{figure}

\begin{figure*}[!tbp]
    \centering
    \includegraphics[width=0.99\textwidth]{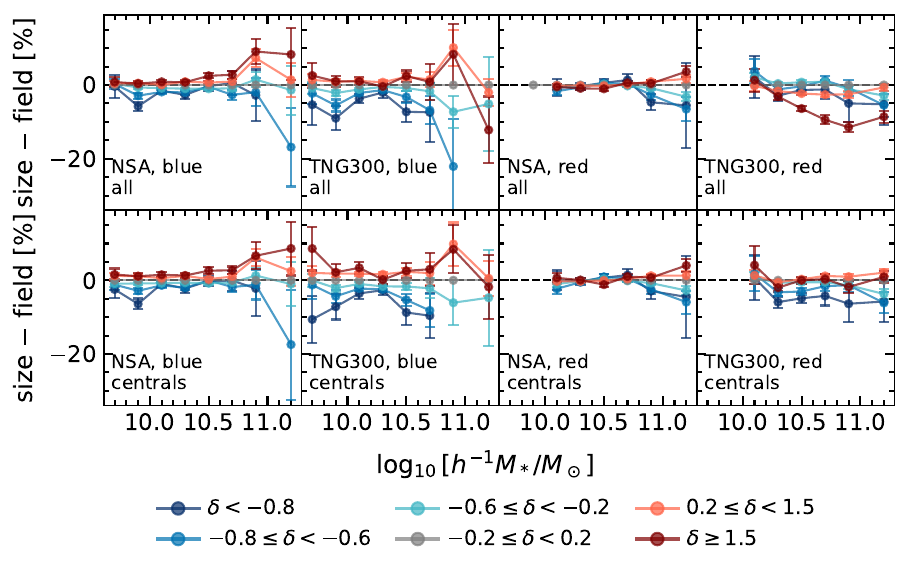}
    \caption{Half-light radius at fixed stellar mass relative to the typical-density galaxies, in six bins of $\delta_{\rm avg}$ as in Figure~\ref{fig:ch7:randsmr}, but split by color. \textit{Left four panels:} the blue galaxies ($g-r<0.7$). \textit{Right four panels:} the red galaxies ($g-r\geq0.7$). \textit{Top row:} all galaxies. \textit{Bottom row:} central galaxies. The NSA and \texttt{TNG300} samples are labeled in each panel. Within each panel the residual is measured against the field galaxies of that same color, so the trend is at fixed stellar mass, redshift, and color for the NSA sample and at fixed stellar mass for the \texttt{TNG300}. Here all galaxies have $M_r\leq-20$. Points with bootstrap errors above $15\%$, from the sparsest bins, are suppressed for clarity. The high-mass deficit toward the deepest underdensities is present in both colors.}
    \label{fig:ch7:bycolor}
\end{figure*}

\begin{figure}[!htbp]
    \centering
    \includegraphics[width=\columnwidth]{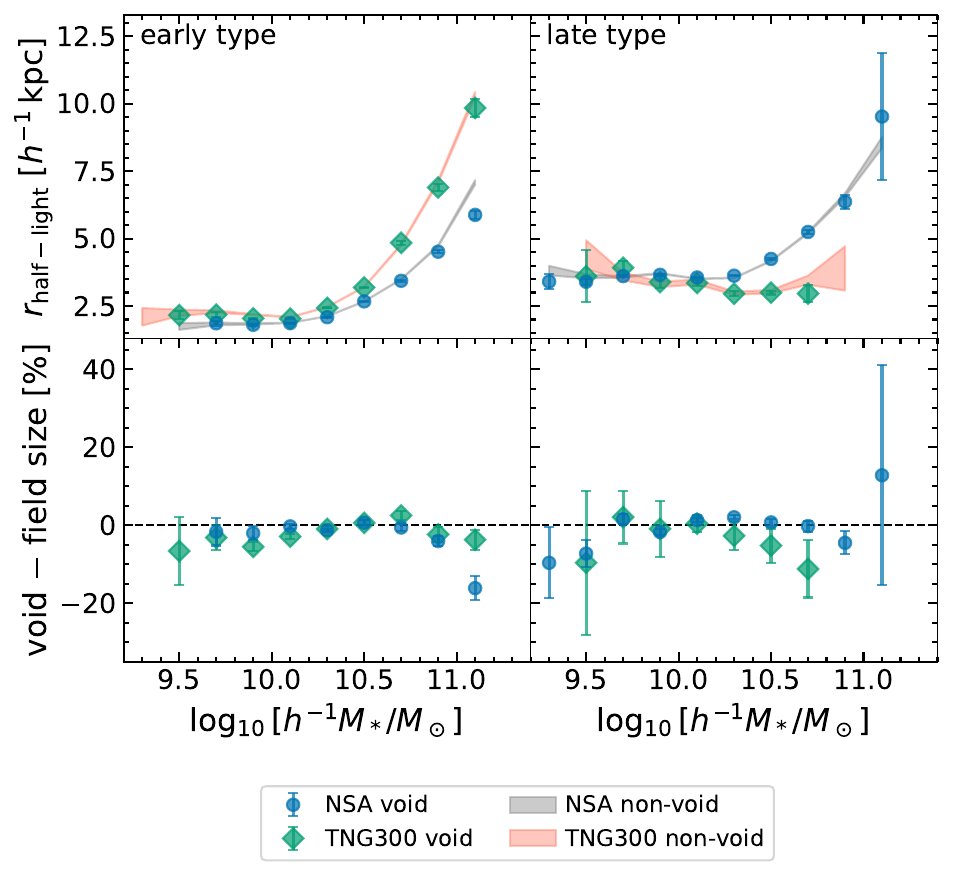}
    \caption{Half-light radius as a function of stellar mass split by morphology. \textit{Left column:} early type. \textit{Right column:} late type. \textit{Top row:} the median half-light radius of the void galaxies, drawn as blue circles for the NSA galaxies and as green diamonds for the \texttt{TNG300} galaxies, against the non-void medians shown as the shaded gray (NSA) and red (\texttt{TNG300}) bands. \textit{Bottom row:} the void-minus-field size residual in percent, measured within each morphology against the field galaxies of that same morphology, at fixed stellar mass, redshift, and color for the NSA sample and at fixed stellar mass for the \texttt{TNG300}, with the NSA error bars from resampling whole voids and the \texttt{TNG300} error bars at the galaxy level. NSA morphologies are the Galaxy Zoo 1 visual classifications of \cite{lintott2011} and the \texttt{TNG300} morphologies are the kinematic cold-disk fractions of \cite{genel2015}. The high-mass void galaxies remain smaller than the fixed-morphology field among the early types in both the data and the simulation. All galaxies have $M_r\leq-20$.}
    \label{fig:ch7:sizemorph}
\end{figure}

The leftmost column of Figure~\ref{fig:ch7:colorcdf} shows $(g-r)$ color ECDFs for all galaxies, for NSA void (circles), NSA non-void (shaded gray), \texttt{TNG300} void (diamonds), and \texttt{TNG300} non-void (shaded red) galaxies at $z=0.0$ (top) and $z=0.1$ (bottom). The underlying color distributions are bimodal. The red peaks of each distribution align at $(g-r)$ colors of $\sim0.75$, but, compared to the blue peaks of the simulated galaxies, the blue peaks of the observed galaxies occur at larger $(g-r)$ values.
Because of this, a wider `green valley' is seen in the simulation data than in the observational data. Lastly, there is a very small population of NSA galaxies with $(g-r)\geq0.8$ ($\sim2\%$ and $\sim5\%$ of the $z=0.0$ and $z=0.1$ populations, respectively) that is absent in \texttt{TNG300}. There are noticeable differences between the color distributions of void and non-void galaxies in both simulations and observations, where, compared to the non-void galaxies, there is an overabundance of void galaxies with $(g-r)<0.65$ and a dearth of galaxies in the red peak.

The medians of these color distributions confirm this picture. In both the observed and simulated cases, the void galaxies tend to have smaller $(g-r)$ colors than their non-void galaxy counterparts. At $z=0.0$, the NSA (\texttt{TNG300}) void galaxies have a median $(g-r)$ color of $0.538\pm0.002$ ($0.452\pm0.001$) and the non-void galaxies have a median $(g-r)$ color of $0.620\pm0.001$ ($0.537\pm0.001$). Similarly, at $z=0.1$, the NSA (\texttt{TNG300}) void galaxies have a median color of $0.657\pm0.001$ ($0.510\pm0.002$) while the non-void galaxies have a median color of $0.699\pm0.001$ ($0.638\pm0.002$).

\subsubsection{Dependency on Stellar Mass}

We now present $(g-r)$ color distributions as a function of stellar mass. We divide each galaxy population into three stellar mass bins that are bounded by $M_*=10^{9.5}$, $10^{10.0}$, $10^{10.5}$, and $10^{11.0}h^{-1}M_\odot$. At $z=0.0$, there are $5{,}455$ ($12{,}355$), $2{,}772$ ($9{,}975$), and $375$ ($2{,}858$) NSA (\texttt{TNG300}) void galaxies and $16{,}228$ ($43{,}378$), $11{,}227$ ($41{,}537$), and $2{,}687$ ($17{,}643$), non-void galaxies in each mass bin. Similarly, at $z=0.1$, there are $3{,}894$ ($3{,}366$), $20{,}623$ ($8{,}240$), and $5{,}315$ ($2{,}325$), NSA (\texttt{TNG300}) void galaxies and $12{,}001$ ($11{,}730$), $92{,}720$ ($37{,}046$), and $35{,}313$ ($17{,}394$), non-void galaxies in each mass bin.

The three rightmost columns of Figure~\ref{fig:ch7:colorcdf} show the $(g-r)$ ECDFs in these mass bins, for the same four populations, and they reveal distinct trends between these galaxy populations as a function of color. Compared to non-void galaxies across all stellar mass bins, void galaxies are preferentially bluer. In the lowest stellar mass bin, this appears as an excess of void galaxies around the blue peak that is centered at $(g-r)\sim0.35-0.4$, while, in the second and third stellar mass bin, this manifests as an excess of galaxies at $(g-r)\sim0.6$ for the NSA galaxies and $(g-r)\sim0.45$ for the \texttt{TNG300} galaxies. Across all mass bins, there is little difference between the probability densities of the red peaks, which are comparable for both the void and non-void galaxies.

There are also apparent differences between the NSA and \texttt{TNG300} samples. For instance, compared to the observed galaxies in the lowest stellar mass bin, the simulated galaxies show a wider distribution of colors, and they have peaks that are $\sim0.05$ magnitudes bluer. In the second stellar mass bin, there are more \texttt{TNG300} void galaxies with $(g-r)>0.6$ than there are in the lowest stellar mass bin, but there is still a sizable population of simulated and observed galaxies with $(g-r)<0.6$. Compared to the observed galaxies in this stellar mass bin, the simulated distributions have blue peaks that are $\sim0.1$ magnitudes bluer and red peaks that are $\sim0.05$ magnitudes higher. 
Finally, the observed galaxies show a wider distribution in this bin, with $\sim40\%$ of the observed void galaxies and $\sim20\%$ of the simulated void galaxies having $(g-r)<0.7$. 

The medians of these color distributions, which appear as the $50\%$ crossings of the empirical cumulative distribution functions in Figure~\ref{fig:ch7:colorcdf}, are lower for the void galaxies than for the non-void galaxies in most cases, and they are also lower for the simulated galaxies than for the observed galaxies. The second stellar mass bin shows the largest discrepancy between void and non-void galaxy populations, where at $z=0.1$, $\Delta(g-r)_{\rm NSA}\equiv (g-r)_{\rm NSA,non-void}-(g-r)_{\rm NSA,void}=0.028\pm0.002$. These differences are more pronounced in the simulated sample, where in the same mass and redshift bin, $\Delta(g-r)_{\rm TNG300}=0.067\pm0.003$.  %From this, it is clear that, compared to the NSA sample, there is an abundance of the bluest galaxies in all stellar mass bins. For example, comparing the void galaxies at $z=0.0$, we find $\Delta(g-r)\equiv (g-r)_{\rm NSA}-(g-r)_{\rm \texttt{TNG300}}=0.053\pm0.007$. and it overproduced the reddest galaxies in the highest stellar mass bin $(\Delta(g-r)=-0.053\pm0.005$ at $z=0.1$). Compared to the observed galaxies in the second stellar mass bins, the observed galaxies are slightly redder at $z=0.0$ ($\Delta(g-r)=-0.020\pm0.009$) and slightly bluer at $z=0.1$ ($\Delta(g-r)=0.018\pm0.007$). %As such, there is excellent agreement between the population statistics of these distributions (down to $\sim0.05$ magnitudes), but there are still noticeable differences in the shapes of the $P(g-r)$ distributions.

We now test whether the size deficit survives in fixed color bins. Figure~\ref{fig:ch7:sizecolor} recomputes the size-mass relation separately for the blue and the red galaxies, and the high-mass deficit persists in both. It is about $12\pm11\%$ among the blue galaxies, where the high-mass sample is small, and $10\pm4\%$ among the red galaxies, consistent with the full-sample deficit. In terms of our counts-in-cells comparison, splitting the residual by color confirms that the deficit is not a color-mixture effect. Here, only blue galaxies are used when defining the field trend in the left column and vice versa for the right column. Each panel of Figure~\ref{fig:ch7:bycolor} then repeats this process for our counts-in-cells experiment, where each panel references its color subsample to the typical-density bin through Equation~\ref{eq:ch7:resid}, making that bin zero by construction. The high-mass deficit toward the deepest underdensities appears in both the blue and the red galaxies, in the full sample and in the centrals, albeit at less significance due to the weak counting statistics.

\begin{figure*}[!tbp]
    \centering
    \includegraphics[width=0.99\textwidth]{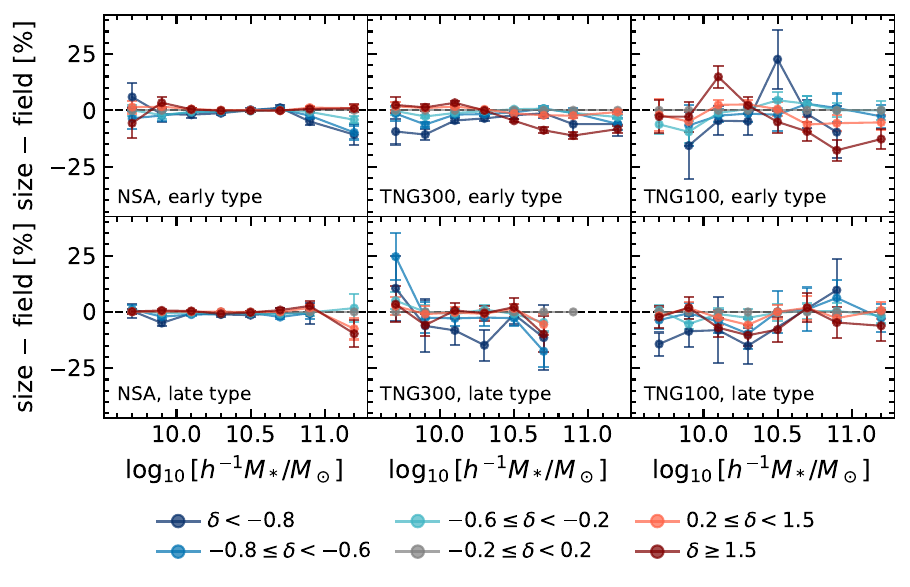}
    \caption{Half-light radius at fixed stellar mass relative to the typical-density galaxies, in six bins of $\delta_{\rm avg}$ as in Figure~\ref{fig:ch7:randsmr}, but split by morphology. \textit{Top row:} early type. \textit{Bottom row:} late type. In each row the columns are the NSA sample (left), the \texttt{TNG300} (center), and the \texttt{TNG100} (right). Within each morphology panel the residual is measured against the field galaxies of that same morphology, at fixed stellar mass, redshift, and color for the NSA sample and at fixed stellar mass for the \texttt{TNG300} and \texttt{TNG100}. NSA morphologies are the Galaxy Zoo 1 visual classifications of \cite{lintott2011}, the \texttt{TNG300} morphologies are the kinematic cold-disk fractions of \cite{genel2015}, and the \texttt{TNG100} morphologies are the deep-learning photometric classifications of \cite{huertascompany2019}. All galaxies have $M_r\leq-20$. The high-mass deficit toward the deepest underdensities is present for the early-type galaxies in all three samples. The \texttt{TNG100} error bars are larger because its smaller box contains fewer deep underdensities, and points with bootstrap errors above $15\%$ are suppressed for clarity as in Figure~\ref{fig:ch7:bycolor}.}
    \label{fig:ch7:bymorph}
\end{figure*}

\begin{figure}[!htbp]
    \centering
    \includegraphics[width=0.95\columnwidth]{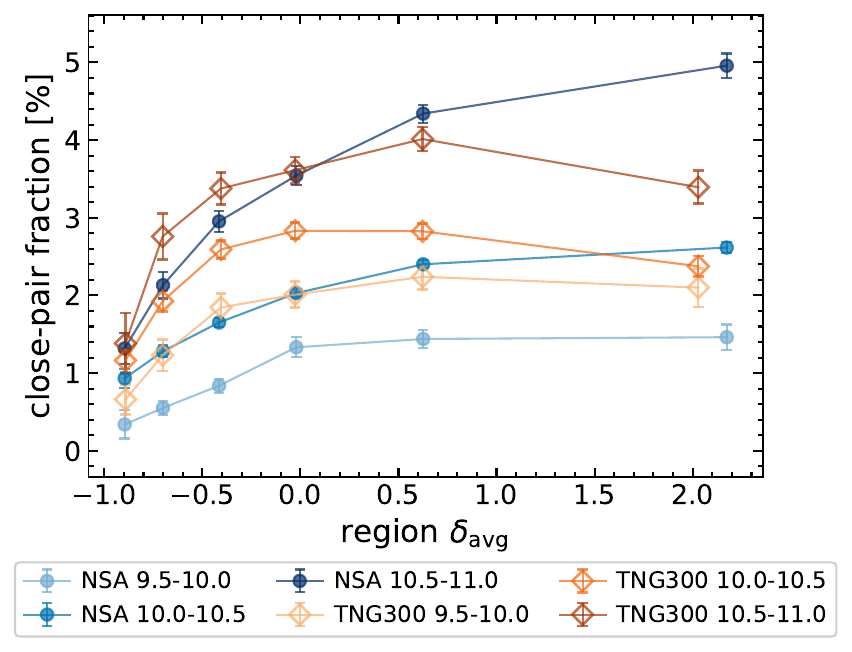}
    \caption{Close-pair fraction as a function of the counts-in-cells underdensity contrast $\delta_{\rm avg}$, in three stellar mass bins, for the NSA sample (filled blue circles) and the \texttt{TNG300} (open orange diamonds). Shades run from light to dark with increasing stellar mass, for the bins $9.5$ to $10.0$, $10.0$ to $10.5$, and $10.5$ to $11.0$ in $\log_{10}(M_\star/M_\odot)$. Here all galaxies have $M_r\leq-20$. A close pair is defined such that there is a companion within $50\,h^{-1}$kpc with stellar mass ratio between $1/4$ and $4$. Error bars are bootstrap estimates. At fixed mass the close-pair fraction falls from typical density toward the deepest underdensities in both samples, so void galaxies are less likely than the field to be in an ongoing close pair.}
    \label{fig:ch7:mergermass}
\end{figure}

\subsection{Morphology}
\label{sec:ch7:morphology}

Our fixed-color test rules out the color mixture as the origin of the deficit. Next, we use the morphological classifications from Section~\ref{sec:ch7:morphmerg} to test whether the deficit persists at fixed morphology (the control that \citealt{Abdullah2026} emphasize).

Figure~\ref{fig:ch7:sizemorph} shows the size residual of our void vs.\ non-void sample split by morphological type. From this figure, it is clear that the high-mass deficit persists within the early-type galaxies, where the NSA void galaxies are about $14\%$ smaller than the early-type field galaxies at $3.8\sigma$, while the late-types are too rare at high mass to constrain. Because the most massive galaxies are overwhelmingly early-type, the deficit is not attributable to the morphological mixture.

Figure~\ref{fig:ch7:bymorph} shows the same test using $\delta_{\rm avg}$. From this figure, the early-type, high-mass deficit toward the deepest underdensities is present in the NSA sample, the \texttt{TNG300} simulation, and the \texttt{TNG100} simulation, reaching about $10\%$ in the deepest bin, while the late types show no trend. As described in Section~\ref{sec:ch7:morphmerg}, the \texttt{TNG100} deep-learning split matches the Galaxy Zoo balance, whereas the \texttt{TNG300} kinematic split is far coarser, and the small \texttt{TNG100} box limits it to this counts-in-cells test. Despite its noisier deepest bins, the early-type trend agrees across all three samples.

\subsection{Recent mergers}
\label{sec:ch7:mergers}

The compactness of massive void galaxies suggests suppressed late-time merger growth, so next we use the close-pair fractions of Section~\ref{sec:ch7:morphmerg} to test whether voids host fewer ongoing mergers. Figure~\ref{fig:ch7:mergermass} shows the close-pair fraction as a function of $\delta_{\rm avg}$ in three stellar mass bins. In every bin and in both samples the fraction decreases from typical density toward the deepest underdensities, and in the NSA sample the decline steepens with mass, mirroring the concentration of the size deficit at the high-mass end. Integrated over the samples, it is $1.4\%$ against $2.8\%$ in the NSA sample and $1.9\%$ against $3.0\%$ in the \texttt{TNG300} sample.

At fixed stellar mass the NSA deficit is significant in each bin, $0.60\%$ against $1.46\%$, $1.32\%$ against $2.35\%$, and $2.00\%$ against $4.27\%$ from the lowest to the highest mass bin, at $4.3\sigma$, $9.6\sigma$, and $8.2\sigma$, respectively. The \texttt{TNG300} shows the same ordering, significant in its two higher mass bins at $5.7\sigma$ and $3.7\sigma$. Although small number statistics prevent us from analyzing the size-mass relation of galaxies undergoing mergers, these results play into our analysis of the size-mass deficit that we observe (e.g., \S~\ref{sec:ch7:disc:mergers}) as it is evidence for a suppression of merger-driven growth in the most massive void galaxies. We note that this deficit concerns the present-day ongoing merger rate, not the cumulative growth budget, and \cite{rodriguesmedrano2024} find that void galaxies undergo a comparable total number of major mergers, occurring later in cosmic time. Still, these results give a glimpse into how mass assembly is occurring on average across populations.

Lastly, when using the Galaxy Zoo 1 visual merger flag instead of our distance- and kinematics-based selection, the same trends are observed. That is, the close-pair fraction increases almost monotonically with stellar mass and large-scale structure density, the latter only being overturned in the densest \texttt{TNG300} regions where the more dynamic cluster environments seem to be slowing merger rates at the redshifts we observe. We note that the NSA fractions are mild lower limits, since SDSS fiber collisions miss the closest pairs, but the void-versus-field trend is robust to this. Only a few percent of galaxies are in a close pair, and removing them leaves the size deficit intact, so the deficit belongs to the bulk population rather than to a merging minority.
\section{Discussion}
\label{sec:ch7:discussion}

\subsection{Environmental trends and delayed evolution}
\label{sec:ch7:disc:trends}

In this work, we have compared the global properties of void and non-void galaxies in the SDSS DR7 and the \texttt{TNG300} simulation, focusing primarily on how large-scale environment shapes galaxy populations. Across luminosity, stellar mass, color, and size, consistent differences emerge between galaxies residing in voids and those in denser regions.

Void galaxies are systematically less massive and bluer than their non-void counterparts. Although the differences in median stellar mass are modest ($\sim0.06$ dex at $z=0.1$ and $\sim0.14$ dex in the deeper $z\leq0.05$ sample) in the NSA catalog, they are persistent across redshift bins and are slightly more pronounced in \texttt{TNG300}. The stellar mass distributions show a relative excess of void galaxies at intermediate masses ($10^{9.5}$ to $10^{10}h^{-1}M_\odot$) and a deficit at the highest masses, indicating that the void environment suppresses the formation or survival of the most massive systems. The color distributions also demonstrate an overabundance of blue galaxies in voids and a relative dearth of galaxies in the red peak. Moreover, the most massive NSA void galaxies have smaller $r$-band half-light radii than their non-void counterparts, a result we quantify in Section~\ref{sec:ch7:density} and that corroborates the findings of \cite{Perez2025}. These trends support the picture in which void galaxies follow delayed evolutionary pathways compared to galaxies in denser environments.

\subsection{The size deficit tracks the underdensity contrast}
\label{sec:ch7:disc:contrast}

The high-mass size deficit is a property of the underdensity contrast rather than of the void label. It deepens toward the emptiest environments and washes out toward typical density (Section~\ref{sec:ch7:density}), running from consistent with zero near the mean density to about a ten percent deficit in the deepest cells. A simple binary ``void-versus-field'' split averages over the whole range of interior densities that voids span, so it dilutes a signal that lives only at the extreme and can hide it entirely when the void sample is shallow. Direct control over the contrast is therefore essential for environmental studies of the size-mass relation, and it is what lets otherwise discrepant measurements be placed on a common axis and compared.

Compared to distances to other cosmic web structures, the underdensity contrast is also a cleaner environmental axis. Voids follow a nearly universal, self-similar density profile \citep{sheth,hamaus2014}, so a fixed distance from a void center maps to a consistent $\delta_{\rm avg}$ from one void to the next. Filaments instead range from short dense bridges between massive halos to long tenuous strands threading the voids \citep{galarraga2020}, and the same distance from a filament spine can therefore correspond to very different local densities. A metric built on distance to a filament or to the nearest cluster then bins together galaxies of very different true density, which blurs any real trend and is a plausible reason that analyses relying on such proxies recover weaker or null environmental effects. Measuring $\delta_{\rm avg}$ directly avoids this because it assigns every galaxy the same physical quantity (i.e., its local mean density), regardless of which structure happens to be the closest.

\subsection{Reconciling the size-mass literature}
\label{sec:ch7:disc:literature}

There is tension in the literature surrounding a dependence of the size-mass relation on local environment, and there are several recent studies that agree and disagree with our claims (see, e.g., \citealt{Abdullah2026} and references therein.) From the Calar Alto Void Integral-field Treasury (CAVITY) voids, \cite{Perez2025} find that void galaxies are smaller at fixed mass (an effect that is greatest for early types and the most massive systems) in a sample of about $1{,}500$ morphologically classified void galaxies drawn from within $0.8$ effective radii of each void center. \cite{Conrado2024} instead find void galaxies to be slightly larger overall, by about $0.04$ dex, from integral-field spectroscopy of $127$ void galaxies in the same voids, and they attribute this to a lower stellar-mass surface density rather than to a size excess at fixed mass. \cite{Porter2023} and \cite{Abdullah2026} find no size dependence at fixed mass and type, the former in $502$ Galaxy And Mass Assembly (GAMA) void galaxies selected by their distance from filaments and the latter in about $148{,}500$ SDSS galaxies split between rich clusters and isolating cylinders. The opposite signs recovered from the same CAVITY voids, with SDSS Petrosian radii in one case and integral-field radii in the other, show how sensitive the result is to the size estimator. In our own measurement, the void galaxies are smaller than the field only at the high-mass end, like \cite{Perez2025}, and at lower masses an excess as small as that of \cite{Conrado2024} would lie within our uncertainties.

The disagreement tracks how each study defines its environment, and the studies that recover a size difference are exactly those that reach genuinely deep underdensities. We select voids with a spherical void finder, \cite{Perez2025} keep only the galaxies within $0.8$ effective radii of a void center, while our earlier \texttt{TNG300} study \citep{Curtis2026} cuts directly on $\delta_{\rm avg}$. The studies that report no dependence instead mostly rely on a weaker proxy for environment. \cite{zhangyang2019} and \cite{Porter2023} label the cosmic web from the tidal tensor or from a spanning tree of galaxies, while \cite{Liao2026} rank galaxies by their distance to the nearest cluster in bins as wide as $30$ to $50$ Mpc. Each of these proxies mixes in regions of only modest underdensity and dilutes a signal that lives only in the deepest voids. The same pattern runs through Table~3 of \cite{Abdullah2026}, where every study that selects a genuine underdensity recovers a dependence, and the lone tabulated null \citep{zhangyang2019} classifies the web by tidal-tensor eigenvalues rather than by a measured density contrast.

The isolation criterion of \cite{Abdullah2026}, a cylinder of radius $2\,h^{-1}$Mpc and velocity depth $3000$ km/s holding at most eight galaxies, does select genuine underdensities, even though they never report the contrast, so their null requires a different explanation. Independent testing reveals that, when we relabel our own galaxies with a cylinder of the same form, the most isolated reach comparable underdensities and recover the same high-mass deficit. Because their sample is itself volume limited, the null reflects neither the environment nor incompleteness. It more plausibly stems from their control on galaxy type, which they hold fixed through specific star formation rate, color, and bulge-to-total ratio together, the last of which can absorb the compactness of the massive early types that exhibit the deficit. While it is true that a color or morphology mixture can masquerade as a size offset, and void galaxies are indeed late-type rich \citep{Argudo2024}, our deficit survives separately at both fixed color and fixed morphology.

\subsection{Suppressed merger-driven growth}
\label{sec:ch7:disc:mergers}

A natural explanation for the suppression of merger-driven growth is the two-phase assembly of massive galaxies, which build their centers early through in-situ star formation and grow at late times by accreting an extended envelope through dry minor mergers \citep{naab2009,oser2010,vandokkum2010}. Because a minor merger deposits its stars in the outskirts of the larger galaxy, the radius of the galaxy grows faster than the galaxy's mass, so a compact high-redshift core can roughly quadruple its half-light radius while little more than doubling in stellar mass \citep{naab2009,vandokkum2010}. In the most massive galaxies, the envelope that is built in this way holds the majority of the present-day stars \citep{oser2010}. This accretion tracks the local density, so massive galaxies in dense regions grow larger envelopes while those in the emptiest regions accrete less, or later, and remain compact. Consistent with this, the most massive early types are as much as $20$ to $40\%$ larger at fixed mass in the densest environments than in underdense ones, with correspondingly more extended outer stellar halos, an enhancement that sets in only above $M_*\sim10^{11}h^{-1}M_\odot$ \citep{yoon2017,huang2018}.

The void deficit is the underdense counterpart of this trend. It is confined to the most massive galaxies, where the accreted envelope dominates the light, and it deepens toward the emptiest interiors. The objects that are responsible for this are almost all central galaxies, $98\%$ of which have no more massive companion within $1\,h^{-1}$Mpc, and about three quarters of them have early type morphologies. The deficit therefore sits exactly where merger-driven envelope growth would leave its imprint, and the same suppression shifts the mass function to lower masses in deeper voids. The mass scale above which our deficit appears is the same scale above which the dense-environment size enhancement sets in \citep{yoon2017}, as expected if a single quantity, the accreted envelope, responds in opposite directions to the high merger rate in clusters and the low merger rate in voids.

Our close-pair measurement supports this picture since void galaxies are less likely to be in an ongoing merger in the present day. Indeed, \cite{rodriguesmedrano2024} find that void galaxies in the same simulation undergo a comparable total number of mergers but experience them later in cosmic time, so they accrete a larger fraction of their stellar mass in the most recent few Gyr, an effect which is strongest for low mass galaxies. That they present this delayed accretion is a challenge to the notion that void galaxies grow via few mergers, so the low present-day close-pair rate we measure need not imply a smaller cumulative merger budget. For the massive galaxies that carry our deficit, then, a truly distinct assembly history vs.\ an assembly rate that is merely delayed remain difficult to separate. Explicit stellar population synthesis modeling (e.g., \citealt{conroy2009,Conroy2013}) of void galaxies could reveal individual galaxy star formation histories (e.g., \citealt{leja2017}) that could reveal bursts and clues to how void galaxies have assembled their stellar mass over cosmic time. Indeed, \cite{cavity1} did just this for a relatively small sample of a few thousand void, field, and cluster galaxies, quantifying the fact that void galaxies appear to assemble their stellar mass slowly, oftentimes being delayed by up to $1$Gyr compared to those in denser environments. Although beyond the scope of this study, extending this analysis to an entire population of galaxies could further elucidate how the assembly history of a galaxy scales with the density of its local environment.

\subsection{Comparison with the \texttt{TNG300} simulation}
\label{sec:ch7:disc:sims}

The luminosity functions reinforce the interpretation above. Non-void galaxies exhibit systematically brighter characteristic magnitudes than void galaxies in both the observed and simulated samples, indicating that the most luminous systems preferentially reside in denser regions of the cosmic web. While the faint-end slopes do not show consistent differences between void and non-void galaxies in the NSA data (in agreement with \citealt{Hoyle2005}, who found only the characteristic magnitude shifted fainter), the \texttt{TNG300} produces shallower slopes for void galaxies, suggesting that environmental effects on low-luminosity systems may depend sensitively on feedback prescriptions or resolution. We caution that the $M_r\leq-20$ selection reaches only about half a magnitude fainter than $M_*$, so the faint-end slopes are weakly constrained and comparisons of $\alpha$ with deeper studies such as \cite{Hoyle2005} should be read with this in mind. %These environmental differences are present in both redshift bins, which implies that the impact of large-scale structure on galaxy evolution is already established by $z\sim0.1$.

When comparing the observations with the results of the simulation, the void effective radii and average underdensity contrasts agree well between the NSA sample and the \texttt{TNG300}, indicating that the simulation reproduces the global structure of underdense regions. Galaxy stellar mass distributions and overall luminosity trends are also broadly consistent. However, secondary discrepancies remain. Simulated void galaxies are systematically bluer than observed void galaxies, with a more pronounced green valley, and the faint-end slopes of the simulated luminosity functions are flatter. Following \cite{illustris1}, the broader \texttt{TNG} galaxy population shows generally good agreement with observed $g-r$ colors. Although both catalogs are restricted to $M_r\leq-20$ in order to match the sample selection of \cite{douglass2023}, this is a luminosity cut rather than a direct constraint on stellar mass, star formation, or dust content. As a result, the NSA and \texttt{TNG} samples do not populate color space identically, and the observed distributions are additionally broadened by real photometric and k-correction uncertainties, while the simulated colors remain comparatively cleaner, producing a more pronounced blue peak in \texttt{TNG}. Moreover, the simulated and observed void galaxies follow broadly similar size-mass relations, with the \texttt{TNG300} void galaxies being modestly smaller in the median. These differences suggest that, while the large-scale environmental trends are robust, the detailed balance of star formation and quenching in low-density regions may still be imperfectly modeled.

Taken together, our results reinforce the conclusion that environment is the primary driver of the trends we find here. That is, void galaxies are, on average, less evolved systems than galaxies in denser regions. The fact that \texttt{TNG300} reproduces these trends to first order indicates that current galaxy formation models capture much of the physics governing galaxy evolution in underdense regions. At the same time, the residual discrepancies in color distributions, faint-end slopes, and sizes highlight areas where improvements in feedback modeling or resolution may be required.

\section{Conclusions}
\label{sec:ch7:conclusions}

We have compared the luminosity functions, stellar mass functions, colors, and mass-size relations of void and non-void galaxies in the complete SDSS DR7 with an identically selected \texttt{TNG300} sample, tying every comparison directly to the average underdensity contrast $\delta_{\rm avg}$ of each host void. Our central result is that the most massive void galaxies are more compact than the mass-matched field, a deficit the simulation reproduces and that deepens toward the emptiest interiors. Our main conclusions are as follows:

\begin{itemize}

\item[$\bullet$] Direct control over the large-scale density contrast is as important to environmental galaxy studies as the selection of the sample itself. The high-mass size deficit depends steeply on the underdensity contrast and washes out toward typical density. A classification that admits only modest underdensity therefore recovers a diluted signal, a natural explanation for the range of conclusions in the recent size-mass literature \citep{Perez2025,Conrado2024,Porter2023,Abdullah2026,Liao2026}.

\item[$\bullet$] Spherical void finders such as \texttt{VoidFinder} suit this measurement because they identify the largest regions that satisfy a fixed underdensity criterion. Our voids have a median $\delta_{\rm avg}\simeq-0.71$, with interiors reaching the $\delta_{\rm avg}\lesssim-0.8$ shell-crossing regime where the deficit is strongest. These largest, most underdense structures are the cleanest laboratories for studying galaxy evolution with minimal dense-environment processing and merit detailed follow-up.

\item[$\bullet$] The most massive void galaxies are more compact than the field at fixed mass, redshift, and color, by about $11\%$, and the deficit deepens toward the emptiest interiors. This is the structural signature expected when the late-time, merger-driven growth of a massive galaxy's outer envelope is suppressed in a low-density environment \citep{naab2009,oser2010,vandokkum2010,yoon2017,huang2018}. Further, our close-pair fractions independently show that ongoing mergers are rarer in voids than elsewhere. Whether this assembly scenario is distinct for all cosmic time, or is merely delayed, remains an open question, since void galaxies in the same simulation eventually complete a comparable number of mergers as galaxies outside voids \citep{rodriguesmedrano2024}.

\item[$\bullet$] The high-mass size deficit is still recovered whether the environment is defined by the \texttt{VoidFinder} voids or by the counts-in-cells contrast, and it persists for blue and red subsamples and for central galaxies alone. It is therefore a genuine structural property of massive galaxies in deep underdensities, not an artifact of the environment definition, the population mixture, or the tidal stripping of satellites.

\end{itemize}

Bulk population studies of this kind remain among the most effective ways to isolate the role of large-scale structure in galaxy evolution. The Dark Energy Spectroscopic Instrument \citep{desi}, which has completed its initial planned survey and continues to observe, \textit{Euclid} \citep{Euclid}, the Hobby-Eberly Telescope Dark Energy Experiment \citep{gebhardt2021}, and the High Latitude Wide Area Survey of the Nancy Grace Roman Space Telescope \citep{HLWAS,verza2025} will map the cosmic web over far larger volumes, but, until their void catalogs mature, the SDSS remains one of the most complete maps available for this purpose. The void population as a whole is consistent with a delayed evolutionary history, while the structural distinctiveness of its most massive members offers a clear target for the deeper void samples these surveys will reveal.

\section*{Acknowledgements}
We are grateful to the anonymous reviewer for helpful comments and suggestions that improved the manuscript. This work was partially supported by National Science Foundation grant AST-2009397. O.C. acknowledges support as a Penn State Extraterrestrial Intelligence Center Postdoctoral Fellow, which is funded through a private donation made by The Ultraintelligence Foundation. The IllustrisTNG simulations were undertaken with compute time awarded by the Gauss Centre for Supercomputing (GCS) under GCS Large-Scale Projects GCS-ILLU and GCS-DWAR on the GCS share of the supercomputer Hazel Hen at the High Performance Computing Center Stuttgart (HLRS), as well as on the machines of the Max Planck Computing and Data Facility (MPCDF) in Garching, Germany. In addition, we are pleased to acknowledge that the computational work reported on in this paper was performed on the Shared Computing Cluster, which is administered by Boston University’s Research Computing Services. BM acknowledges support from Northeastern University's Future Faculty Postdoctoral Fellowship program.

\bibliography{bibliography}{}
\bibliographystyle{aasjournalv7}

\end{document}